\documentclass{aa}

\usepackage[dvipsnames]{xcolor}
\usepackage{tikz,hyperref}

\definecolor{lime}{HTML}{A6CE39}
\DeclareRobustCommand{\orcidicon}{
	\begin{tikzpicture}
	\draw[lime, fill=lime] (0,0) 
	circle [radius=0.13] 
	node[white] {{\fontfamily{qag}\selectfont \tiny ID}};
	\draw[white, fill=white] (-0.0625,0.095) 
	circle [radius=0.007];
	\end{tikzpicture}
	\hspace{-2mm}
}

\foreach \x in {A, ..., Z}{\expandafter\xdef\csname orcid\x\endcsname{\noexpand\href{https://orcid.org/\csname orcidauthor\x\endcsname}
			{\noexpand\orcidicon}}
}

\usepackage{graphicx}
\usepackage{txfonts}
\usepackage{caption}
\usepackage{subcaption}

\begin{document} 

\title{Nova shells in 3D}

\subtitle{I. A catalog of integral field spectroscopic observations}

\author{
Rodríguez-González, J. B.\inst{1}
  \and
Santamaría, E.\inst{2,3}
  \and
M.~A.~Guerrero\inst{1}
  \and
J.~A.~Toal\'{a}\inst{2}
          }

\institute{Instituto de Astrof\'{i}sica de Andaluc\'{i}a, IAA-CSIC, Glorieta de la Astronom\'{i}a S/N, Granada 18008, Spain
         \and
             Instituto de Radioastronom\'{i}a y Astrof\'{i}sica, Universidad Nacional Aut\'{o}noma de M\'{e}xico, Morelia 58089, Michoac\'{a}n, Mexico 
             \and
             Instituto de Astronomía y Meteorología, Dpto. de Física, CUCEI, Universidad de Guadalajara, Guadalajara, Jalisco, Mexico  
             \\
             \email{janis@iaa.es}
             }

   \date{\today}

\abstract{Classical novae eject material during thermonuclear eruptions on the surface of accreting white dwarfs, forming expanding shells (the so-called nova remnants or nova shells). 
The physical structure and kinematics of nova shells shed light onto the eruption mechanism and binary interaction. 
Nova shells are typically characterized using direct imaging and spectroscopy, but their intrinsic three-dimensional structure remains poorly constrained.
}{
The aim of this work is to present the first three-dimensional catalog of nova shells based on integral field spectroscopic observations, which allows investigating the diversity of their physical structures beyond projection effects.
}{
We have compiled integral-field spectroscopic observations of 19 nova shells obtained with MUSE at VLT, complemented with new observations obtained with MEGARA at GTC and WEAVE at WHT. 
Three-dimensional visualizations of emission line of interest have been constructed. 
}{
The three-dimensional visualizations of nova shells reveal a wide variety of structures, including ellipsoidal or spherical (round) shells, ring-like and bipolar features, and clumpy or irregular components. 
Many of these structures are not evident in two-dimensional projections alone. 
Approximately 60\% of the sample exhibits predominantly ellipsoidal or spherical shapes, while about 30\% shows ring-like structures or are bipolar. 
For several objects, strong line-dependent morphological differences are observed, with the [O\,{\sc iii}] or [N~{\sc ii}] emission lines tracing distinct kinematic components to those in H$\alpha$.
}{
Compared to imaging studies, integral field spectroscopic observations unveil a larger fraction of bipolar or toroidal nova shells. 
Their physical structure is found to differ considerably among different emission lines, highlighting distinct excitation mechanisms and physical properties within the ejecta. 
The comparison of the expansion velocity on the plane of the sky and along the line of sight of nova shells implies a low value of their intrinsic aspect ratio, consistent the conclusions derived from their projected aspect ratios. 
}

   \keywords{Catalogs --
                Stars: novae, cataclysmic variables, circumstellar matter --
                Techniques: imaging spectroscopy -- ISM: individual objects: DO Aql, V1229 Aql, V1425 Aql, T Aur, V842 Cen, RR Cha, V1974 Cyg, HR Del, BT Mon, RR Pic, CP Pup, CT Ser, FH Ser, V5668 Sgr, XX Tau, NQ Vul, QU Vul
               }

    \maketitle
\section{Introduction}

Classical novae are cataclysmic variable (CV) systems in which a white dwarf (WD) accretes hydrogen-rich material from a companion star, typically a late-type main-sequence or slightly evolved star. 
When critical pressure and temperature values are reached in the accreted envelope, a thermonuclear runaway (TNR) ignites \citep{Starrfield1972}, producing a sudden ejection of material and an increase in luminosity by several magnitudes. 
This process has long been studied both observationally and theoretically \citep[e.g.][]{GallagherStarrfield1978,BodeEvans2008}, revealing nova eruptions among the most common and energetic Galactic stellar transients. 
As the ejecta expand at velocities of hundreds to thousands of km\,s$^{-1}$, they undergo complex radiative and dynamical processes, exhibiting rich spectroscopic signatures that evolve from the early optically thick fireball phase to the nebular and coronal stages \citep{Warner1995,Shore2012}.

The physical structure and kinematics of the nova ejecta conceal essential information to constrain the physics of mass loss, binary interaction, and eruption mechanisms. 
Early theoretical work predicted strong deviations from spherical symmetry, with bipolar lobes, equatorial rings, and clumpy substructures arising from anisotropic mass ejection caused by binarity and stellar rotation \citep{Livio1990,Lloyd1997,Porter1998}.

Observational studies have since investigated this theoretical scenario. 
Several major catalogs compiled photometric, spectroscopic, and basic morphological properties of Galactic novae \citep{PayneGaposchkin1964,Duerbeck1987,Warner1995}. 
More recent compilations have incorporated multiwavelength observations and derived physical parameters \citep[see chapters 7 to 10 of][]{BodeEvans2008,WoudtRibeiro2014}. 
Dedicated surveys of resolved nova shells based on narrow-band imaging \citep{Slavin1995,DownesDuerbeck2000} have provided valuable information on their morphology. 
Recently, \citet{Santamaria2025a} presented the most extensive morphological catalog of resolved nova shells to date, based on deep narrow-band imaging. 
This work shows elliptical or round morphologies to be predominant, with a tiny fraction of nova shells displaying bipolar and axisymmetric structures. 

Since the morphology of a nova shell informs only of the emissivity-weighted projection of its physical structure onto the plane of the sky, complementary progress can be achieved adding spectroscopic information, which provide direct access to the kinematic of nova ejecta. 
Long-slit spectroscopy of nova shells\footnote{The high expansion velocity of nova shells allows investigating their kinematics using data with relatively low spectral resolution, $R \geq2000$.} can provide position-velocity maps that, in conjunction with imaging, can be used to obtain three-dimensional (3D) models of the nova shell \citep{HarmanOBrien2003,Ribeiro2009}. 
These morpho-kinematic modeling suffer, howeveer, from notable limitations, as combining images and spectra obtained at different epochs, with different spatial resolutions, or using a small number of slit positions leads to ambiguous or degenerate interpretations of nova shell structures \citep{Santamaria2022a}, when different 3D structures result in similar projected morphologies and line profiles.

In contrast, full 3D reconstructions combining spatially resolved spectroscopy and kinematic information unleash the complex internal structure of nova remnants, often exposing deviations from the simplified geometries inferred from 2D modeling alone \citep{Santamaria2022b,Santamaria2024_T_Pyx}. 
These results suggest that discrepancies may exist between morphology inferred from spectro-imaging modeling and the true 3D structure of nova ejecta. 
This highlights the need for systematic 3D analyses to accurately characterize nova shells and to establish a consistent framework for comparing observations with theoretical and hydrodynamical models.
The increasing availability of integral-field spectroscopy (IFS) providing spatially resolved spectra, in conjunction with advanced visualization tools have opened a new window on the study of nova shells. 

These developments provide a powerful framework for linking observed morphologies with the physical processes governing nova eruptions.

In this work, we present the first comprehensive 3D catalog of nova shells based on IFS data. The paper is organized as follows. In Section~\ref{sec:Obs}, we describe the sample and observational data. 
A structural classification scheme, together with a description of the one-dimensional spectra of the objects in the sample, is presented in Section~\ref{sec:Data}. 
The results are discussed in Section~\ref{sec:Discussion}, and the main conclusions are summarized in Sect.~\ref{sec:summary}.

\section{Observations and Data Reduction}
\label{sec:Obs}

\begin{table*}
\centering
\caption{List of nova shells with IFS observations.}
\label{tbl:nova.lst}
\setlength{\tabcolsep}{8pt}
\renewcommand{\arraystretch}{1.1}
\begin{tabular}{lrclccr}
\hline
Source &
\multicolumn{1}{c}{$l,b$} &
Outburst date &
\multicolumn{1}{c}{Observation epoch} &
Telescope &
Instrument &
\multicolumn{1}{c}{Exp.\ time} \\
 & \multicolumn{1}{c}{(deg)} & & & & & \multicolumn{1}{c}{(s)} \\
\hline
DO\,Aql      & 31.70$-$11.80 & 1925 Sep   & ~~~~~~~2023 Aug & VLT & MUSE     & 2186~~~~ \\
V1229\,Aql   & 40.53$-$05.43 & 1970 Apr   & ~~~~~~~2022 Sep & VLT & MUSE     & 3040~~~~ \\
V1425\,Aql   & 33.01$-$03.89 & 1995 Feb   & ~~~~~~~2021 Aug & VLT & MUSE     & 1200~~~~ \\
T\,Aur       & 177.14$-$01.70& 1891 Dec   & ~~~~~~~2022 Oct & GTC & MEGARA   &  480~~~~ \\
V842\,Cen    & 316.57$+$02.45& 1986 Nov   & ~~~~~~~2022 Jan & VLT & MUSE     & 1091~~~~ \\
RR\,Cha      & 304.16$-$19.54& 1953 Apr   & ~~~~~~~2022 Jun & VLT & MUSE     & 2721~~~~ \\
V1974\,Cyg   & 89.13$+$07.81 & 1992 Feb   & ~~~~~~~2021 Aug & GTC & MEGARA   &  600~~~~ \\
J210204\,Cyg & 88.01$+$00.45 & $\dots$--1890 & ~~~~~~~2026 Jan & WHT & WEAVE    &  900~~~~ \\
HR\,Del      & 63.43$-$13.97 & 1967 Jul   & ~~~~~~~2023 Apr & VLT & MUSE     &  337~~~~ \\
BT\,Mon      & 213.86$-$02.62& 1939 Dec   & ~~~~~~~2021 Nov & VLT & MUSE     & 2400~~~~ \\
RR\,Pic      & 272.35$-$25.67& 1925 May   & ~~~~~~~2021 Nov & VLT & MUSE     & 2800~~~~ \\
CP\,Pup      & 252.92$-$00.83& 1942 Nov   & ~~~~~~~2021 Dec & VLT & MUSE     & 1580~~~~ \\
V351\,Pup    & 252.72$-$00.73& 1991 Dec   & ~~~~~~~2021 Dec & VLT & MUSE     & 2800~~~~ \\
CT\,Ser      & 24.48$+$47.56 & 1948 Apr   & ~~~~~~~2023 Aug & VLT & MUSE     & 3000~~~~ \\
FH\,Ser      & 32.90$+$05.78 & 1970 Feb   & ~~~~~~~2023 Jul & VLT & MUSE     & 2160~~~~ \\
V5668\,Sgr   & 05.38$-$09.86 & 2015 Mar   & ~~~~~~~2024 Jul & VLT & MUSE     & 4800~~~~ \\
XX\,Tau      & 187.10$-$11.65& 1927 Nov   & ~~~~~~~2023 Aug & VLT & MUSE     & 1949~~~~ \\
NQ\,Vul      & 55.35$+$01.28 & 1976 Oct   & ~~~~~~~2022 Sep & VLT & MUSE     & 2400~~~~ \\
QU\,Vul      & 68.51$-$06.02 & 1984 Dec   & ~~~~~~~2022 Sep & VLT & MUSE     & 2400~~~~ \\
\hline
\end{tabular}
\end{table*}

We present here a compilation of IFS observations of nova shells obtained from both archival data and proprietary observations. 
The sample consists of a total of 19 nova shells. 
The complete list of objects, together with the observational information, is presented in Table~\ref{tbl:nova.lst}.
The time of the nova explosion has been adopted from \citet{Santamaria2025a}. 

Archival data were retrieved from the ESO\footnote{https://archive.eso.org/cms.html} and GTC science archives. 
These include observations obtained with the Multi Unit Spectroscopic Explorer \citep[MUSE;][]{MUSE2020} at the Very Large Telescope (VLT) and the Multi-Espectrógrafo en GTC de Alta Resolución para Astronomía \citep[MEGARA;][]{megara2018} at the Gran Telescopio Canarias (GTC). 

All novae observed with MUSE were obtained in Wide Field Mode (WFM), except V5668~Sgr, which was observed in Narrow Field Mode (NFM). In WFM, MUSE provides integral-field spectroscopy over a $1\arcmin \times 1\arcmin$ field of view (FoV), with a spatial sampling of $0\farcs2$ and a typical spatial resolution of $0\farcs3$--$0\farcs4$ (FWHM). The instrument covers a broad optical wavelength range from approximately 4650 to 9300\,\AA\ with a spectral resolving power varying from $R \sim 2000$ at 4600 \AA\ to $R \sim 4000$ at 9300 \AA, corresponding to velocity resolutions of roughly 150--75 km\,s$^{-1}$ across the spectral range.

The NFM configuration provides a smaller FoV of $7\farcs5 \times 7\farcs5$ with a spatial sampling of $0\farcs025$ and an adaptive-optics-assisted spatial resolution of $0\farcs03$--$0\farcs05$ (FWHM). 
The spatial resolution is thus significantly enhanced compared to that provided by the WFM mode, whereas the spectral coverage is almost identical.

The MEGARA IFS mode covers a FoV of approximately $12.5 \times 11.3$ arcsec$^{2}$, sampled by 567 hexagonal spaxels with a diameter of $0\farcs62$. 
When used in its high-spectral resolution configuration, the instrument reaches spectral resolving powers up to $R \sim 18700$, corresponding to a velocity resolution of $\sim16$ km s$^{-1}$, over a wavelength range of about 6400–6800 \AA.

In addition, we obtained new observations of IPHASX J210204.7+471015 (hereafter J210204\,Cyg) using the WHT Enhanced Area Velocity Explorer \citep[WEAVE;][]{weave} IFS at the William Herschel Telescope (WHT). 
The observations were carried out using the Large Integral Field Unit (LIFU) mode, which consists of 547 fibres, each $2\farcs6$ in diameter. 
The fibers are arranged over an hexagonal FoV of approximately $78 \times 90$ arcsec$^{2}$ with a filling factor of 0.55. 
The lower-resolution gratings were used, delivering a spectral resolving power of $R \sim 2500$ to both the blue and red arms of the spectrograph. The blue arm covers a wavelength range from 3600 to 5955\,\AA, while the red arm spans 5800 to 9490\,\AA, providing continuous coverage over most of the optical regime.

The MUSE data cubes retrieved from the ESO archive were already processed with the standard ESO pipeline. 
They are fully calibrated in wavelength and flux. 

The raw MEGARA data were reduced following the Data Reduction Cookbook (Universidad Complutense de Madrid; \citealp{Pascual2019}, version released on 2019 July 24). The pipeline performs bias and sky subtraction, flat-field correction, wavelength calibration, spectra tracing, and extraction. Sky subtraction is carried out using 56 ancillary fibres located approximately 2 arcmin from the center of the IFU. Flux calibration was achieved through observations of the spectro-photometric standard HR~7950. The final data cubes were generated using the \texttt{megararss2cube} regularization task, producing cubes of $52 \times 58 \times 4300$ elements with a spatial sampling of 0.215 arcsec.

The WEAVE data were extracted, wavelength and flux calibrated by the WEAVE core processing system operated by CASU in Cambridge \citep{Casu2019}. 

\section{Data analysis}
\label{sec:Data}
\subsection{Two-dimensional emission-line images}

\begin{figure*}
\centering
\includegraphics[width=1.0\linewidth]{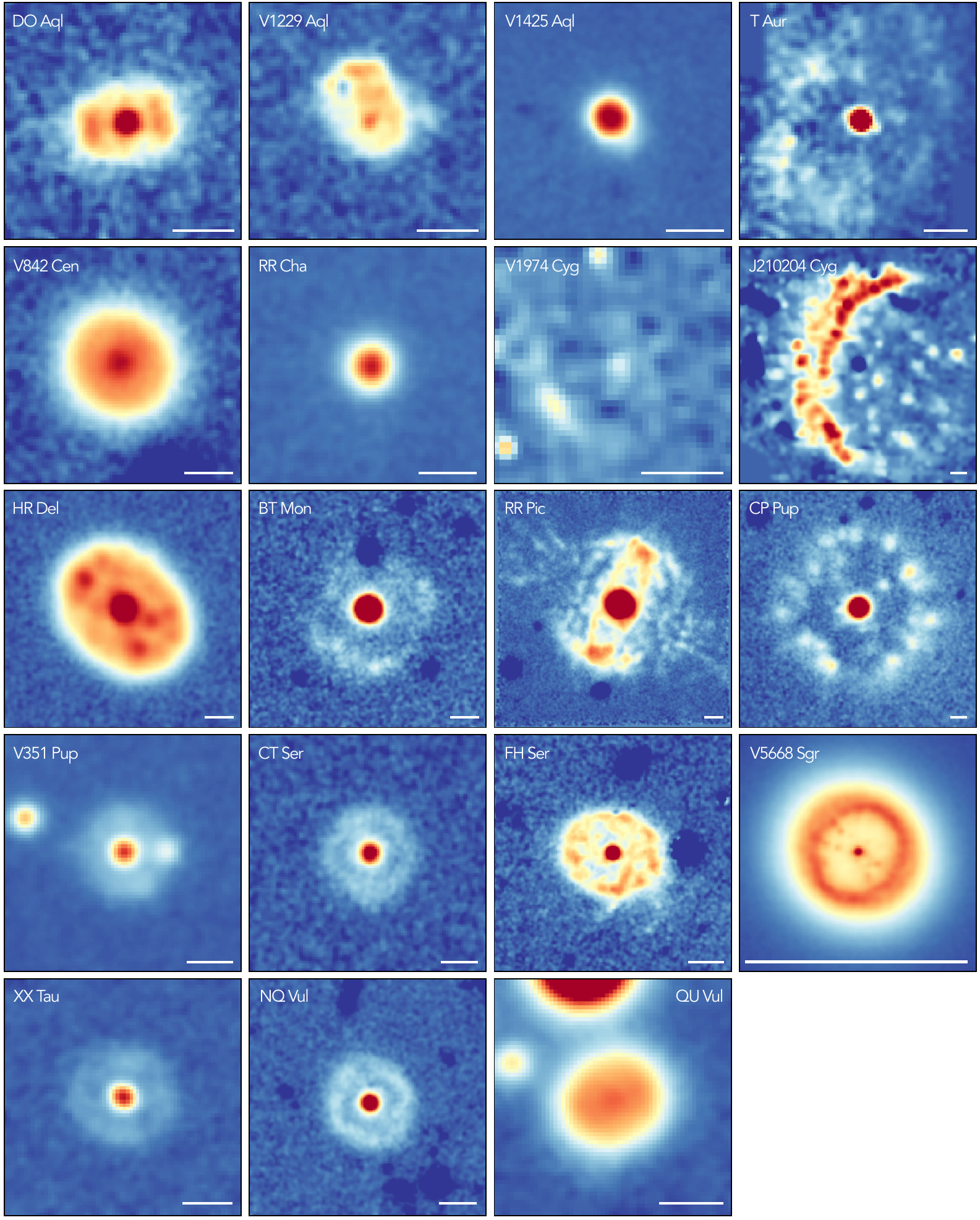}
\caption{
2D emission-line images of the nova shells in the sample listed in Table~\ref{tbl:nova.lst}, extracted from the IFU data cubes. 
All panels show the H$\alpha$ emission, except for J210204\,Cyg, which is displayed in the [N\,{\sc ii}]~$\lambda6584$ emission line. 
The white bar in each panel corresponds to 3~arcsec. 
The colour scaling for each object has been chosen to highlight its morphology. 
North is up and East to the left in all images.
}
\label{FigVibStab}
\end{figure*}

Emission-line images extracted from the IFS data of all nova remnants in our sample are shown in Fig.~\ref{FigVibStab}. 
The scale bar of $3''$ shown in each panel allows an immediate visual comparison of the angular extent of these shells.  
Their angular radii, computed at $3\sigma$ over the background, are listed in Tab.~\ref{tbl:nova.info}.  
When the nova shell is not round, two measurements along the minor and major axes are provided.
The angular radii of the nova shells range from $1\farcs1$ for V5668\,Sgr up to $\simeq35''\times42''$ for J210204~Cyg. 
Using the compilation of distances listed in the third column of Table~\ref{tbl:nova.info}, based on Gaia DR3 parallaxes \citep{BailerJones2021} and refined by \citet{Schaefer2022}\footnote{
The two distance scales seem to be consistent up to 3 kpc \citep[see][]{Santamaria2025a}.
}
the physical radii of the nova shells have been derived and listed in column four of this table.  
Physical radii range from 0.007 pc for V5668\,Sgr up to 0.146 pc for V351\,Pup and 0.131 for XX\,Tau, although the latter have large uncertainties in their distances that directly affect these estimates.  

Columns 5 and 6 in Table~\ref{tbl:nova.info} list the age of the image from the time of the nova explosion and the tangential expansion velocity along the minor and major axes computed as
\begin{equation}
V_{\rm sky} = \frac{r}{t},
\label{eq.vsky}
\end{equation}
where $r$ is the angular radius corrected for seeing effects, and $t$ is the age of the nova since the eruption.
The correction applied to the radius accounts for the point spread function (PSF) and is given by
\begin{equation}
r = r_{\rm obs} - 3\sigma_\star,
\end{equation}
where $r_{\rm obs}$ is the observed angular radius listed in column 4 of Tab.~\ref{tbl:nova.info} and $\sigma_\star$ is the width of the stellar PSF, which is related to the FWHM$_\star$ of stars in the field of view as
\begin{equation}
\sigma_\star = \frac{\mathrm{FWHM}_\star}{2 \sqrt{2 \ln 2}} = \frac{\mathrm{FWHM}_\star}{2.355}.
\end{equation}

\begin{figure}
    \centering
    \includegraphics[width=1.0\linewidth]{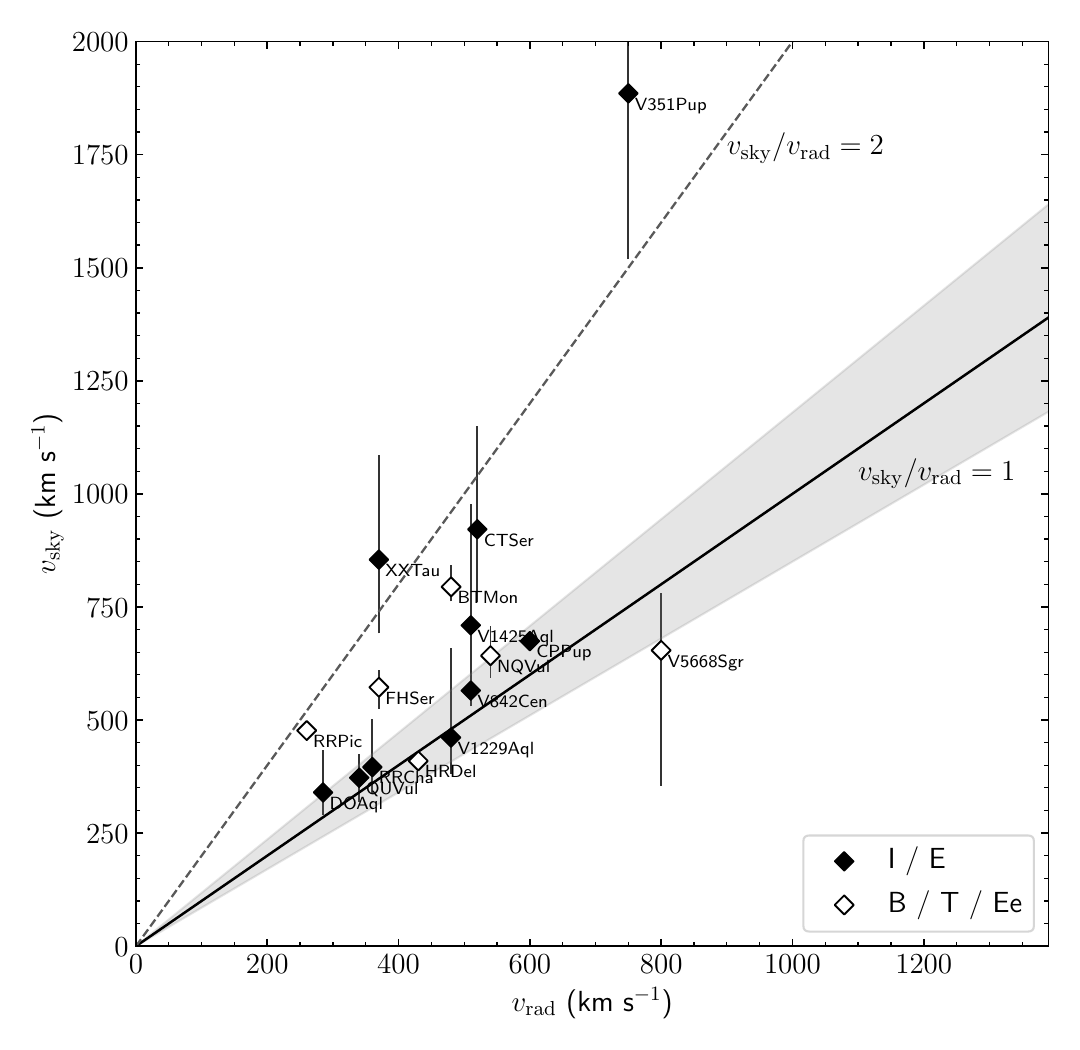}
\caption{
Comparison between expansion velocities projected on the plane of the sky ($V_{\mathrm{sky}}$) and radial expansion velocities ($V_{\mathrm{rad}}$). 
Filled diamonds correspond to Irregular/Elliptical structures (I/E), while open diamonds denote Bipolar, Toroidal or Elliptical with equatorial enhancement (B/T/Ee). The vertical error bars reflect the uncertainties propagated from the distance estimates. 
The solid and dashed lines indicate a velocity ratio $V_{\mathrm{sky}}/V_{\mathrm{rad}}$ of 1 and 2, respectively. The gray shaded area marks the range ($0.8 V_\mathrm{rad} \leq V_\mathrm{sky} \leq 1.18 V_\mathrm{rad}$) reported by \citet{Santamaria2025a}, representing the expected ratio between the plane-of-the-sky and radial velocities.}
    \label{fig:vsky_vrad}
\end{figure}

Tangential expansion velocities range from 220 km~s$^{-1}$, along the minor axis of DO\,Aql, up to 1890 km~s$^{-1}$ along the major axis of V351\,Pup. The radial expansion velocities of the nova shells in the sample are listed in the seventh column of Tab.~\ref{tbl:nova.info} and shown in Fig.~\ref{fig:vsky_vrad} for easy comparison with the tangential velocity component.

\begin{sidewaystable*}[p]
\centering
\caption{Information on the nova shells with IFS observations.} 
\label{tbl:nova.info}
\setlength{\tabcolsep}{6pt}
\renewcommand{\arraystretch}{1.1}
\small
\begin{tabular}{lcccrcrcl}
\hline
\multicolumn{1}{c}{Source} &  
\multicolumn{1}{c}{Angular radius} & 
\multicolumn{1}{c}{Distance} & 
\multicolumn{1}{c}{Physical radius} & 
\multicolumn{1}{c}{Age} & 
\multicolumn{1}{c}{$V_\mathrm{sky}$} & 
\multicolumn{1}{c}{$V_\mathrm{radial}$} &
\multicolumn{1}{c}{Morphology $\rightarrow$ 3D Structure} & 
\multicolumn{1}{c}{H$\alpha$ Profile} \\
\multicolumn{1}{c}{} & 
\multicolumn{1}{c}{(arcsec)} & 
\multicolumn{1}{c}{(kpc)} & 
\multicolumn{1}{c}{(pc)} & 
\multicolumn{1}{c}{(yr)} & 
\multicolumn{2}{c}{~~~~~(km~s$^{-1}$)} &
\multicolumn{1}{c}{} &
\multicolumn{1}{c}{} \\
\hline
DO\,Aql      & $2.3\times3.4$   & $4.4^{+1.2}_{-0.7}$   & $0.049{\bf^{+0.01}_{-0.01}}\times0.072{\bf^{+0.02}_{-0.01}}$ & 97.9  & $220{\bf^{+135}_{-72}}\times450{\bf^{+200}_{-107}}$   & 285             & I $\rightarrow$ E/I       & ~~~~DP \\
V1229\,Aql   & $2.6\times2.7$   & $3.3^{+1.4}_{-0.6}$   & $0.042{^{+0.02}_{-0.01}}\times0.043{^{+0.02}_{-0.01}}$ & 52.4  & 390${^{+347}_{-142}}\times440{^{+334}_{-136}}$   & 480             & Es $\rightarrow$ Ic       & ~~~~DP$^\star$ \\
V1425\,Aql   & 1.7              & $5.7^{+2.2}_{-1.4}$  & 0.047${^{+0.02}_{-0.01}}$              & 26.5  & 410${^{+657}_{-438}}$              & 510             & R $\rightarrow$ E         & ~~~~MP \\
T\,Aur       & $\dots$          & $0.80^{+0.04}_{-0.03}$       & $\dots$            & 130.8 & $\dots$          & $\dots$         & Ect $\rightarrow$ $\dots$ & ~~~~$\dots$ \\
V842\,Cen    & 4.9              & $1.37^{+0.09}_{-0.06}$      & 0.033${^{+0.002}_{-0.002}}$              & 35.2  & 670${^{+59}_{-42}}$              & 510             & Ecft $\rightarrow$ E      & ~~~~FT-ac \\
RR\,Cha      & $2.1\times2.3$   & $6.1^{+1.7}_{-0.9}$   & $0.062{^{+0.02}_{-0.01}}\times0.068{^{+0.02}_{-0.01}}$ & 69.2  & $340{^{+238}_{-134}}\times420{^{+260}_{-147}}$   & 360             & S $\rightarrow$ E/R       & ~~~~SP-a \\
V1974\,Cyg   & $\dots$          & $1.63^{+0.24}_{-0.12}$    & $\dots$            & 29.5  & $\dots$          & $\dots$         & Ect $\rightarrow$ $\dots$ & ~~~~$\dots$ \\
HR\,Del      & $6.9\times8.5$   & $0.872\pm0.014$              & $0.029{\pm0.001}\times0.036{\pm0.001}$ & 55.8  & $410{\pm10}\times530{\pm10}$   & 430             & Ecft $\rightarrow$ B      & ~~~~FT-c$^\star$  \\
BT\,Mon      & $9.6\times10.1$  & $1.50^{+0.09}_{-0.06}$      & $0.070{^{+0.004}_{-0.003}}\times0.073{^{+0.004}_{-0.003}}$ & 81.9  & $720{^{+49}_{-34}}\times770{^{+52}_{-36}}$   & 480             & E $\rightarrow$ Tc        & ~~~~DP-c \\
RR\,Pic      & $20.5\times21.2$ & $0.491^{+0.005}_{-0.006}$         & $0.049{\pm0.001}\times0.050{\pm0.001}$ & 96.5  & $460{^{+5}_{-6}}\times480{^{+5}_{-6}}$   & 260             & Ecte $\rightarrow$ Tc     & ~~~~DP \\
CP\,Pup      & $15.7\times16.9$ & $0.758\pm0.010$              & $0.058{\pm0.001}\times0.062{\pm0.001}$ & 79.1  & $650{\pm10}\times710{\pm10}$   & 600             & Rc $\rightarrow$ ``E''c   & ~~~~FT-c \\
V351\,Pup    & 2.5              & $12.0^{+1.8}_{-2.3}$ $^\dagger$ & 0.146${^{+0.02}_{-0.03}}$              & 30.0  & 1890${^{+725}_{-926}}$             & 750             & Rmx $\rightarrow$ Ec      & ~~~~FT-c\\
CT\,Ser      & 4.7              & $4.0^{+1.0}_{-0.7}$ $^\dagger$   & 0.092${^{+0.02}_{-0.01}}$              & 75.3  &  920${^{+296}_{-207}}$             & 520             & Rc $\rightarrow$ Ec       & ~~~~FT-c \\
FH\,Ser      & $7.4\times8.5$   & $1.08^{+0.07}_{-0.09}$      & $0.039{^{+0.002}_{-0.003}}\times0.045{^{+0.003}_{-0.004}}$ & 63.4  & $500{^{+39}_{-50}}\times590{^{+44}_{-57}}$   & 370             & Ecef                      & ~~~~DP   \\
V5668\,Sgr   & 1.1              & $1.3^{+0.3}_{-0.6}$    & 0.007${\bf^{+0.001}_{-0.003}}$              & 9.3   & 740${\bf^{+140}_{-331}}$              & 800             & Em $\rightarrow$ B        & ~~~~FT-c$^\star$ \\
XX\,Tau      & $3.2\times3.6$   & $7.5^{+2.0}_{-1.4}$ $^\dagger$ & $0.116{\bf^{0.03}_{-0.02}}\times0.131{\bf^{+0.04}_{-0.03}}$ & 95.8  & $780{^{+360}_{-255}}\times930{^{+320}_{-226}}$   & 370             & I $\rightarrow$ Ec        & ~~~~DP \\
NQ\,Vul      & $6.0\times6.6$   & $1.18^{+0.12}_{-0.09}$     & $0.034{^{+0.004}_{-0.003}}\times0.038{^{+0.004}_{-0.003}}$ & 45.9  & $590{^{+74}_{-56}}\times660{^{+82}_{-61}}$   & 340             & Es $\rightarrow$ TBc      & ~~~~FT-c \\
QU\,Vul      & 4.2              & $1.43\pm0.20$            & 0.029${\pm0.004}$             & 37.8  & 530${\pm107}$            & 540             & Efmx $\rightarrow$ E      & ~~~~FT-c \\
\hline
\end{tabular}
\tablefoot{The last column provides the H$\alpha$ spectral-line profile classification: single-peaked (SP), double-peaked (DP), multi-peaked (MP), or flat-top (FT), with secondary descriptors indicating asymmetric (a) or castellated (c) profiles.\\
\tablefoottext{$\dagger$}{
The distances to V351\,Pup, CT\,Ser and XX\,Tau imply $V_\mathrm{sky}$ much larger than $V_\mathrm{rad}$.  
If both velocities were similar, placing them within the gray-shaded region of Fig.~\ref{fig:vsky_vrad}, their distances would need to be reduced to $4.8\pm0.8$ kpc for V351\,Pup, $2.3\pm0.4$ kpc for CT\,Ser, and $3.3\pm0.8$ kpc for XX\,Tau. 
}
\tablefoottext{$^\star$}{
The H$\alpha$ line is blended or contaminated by the [N~{\sc ii}] emission lines.
}
\\
J210204 Cyg is not included in this classification because no H$\alpha$ emission is detected in its spectrum.
}
\end{sidewaystable*}

In several cases, the central binary system is visible as a bright point source surrounded by diffuse nebular emission, while in others the shell emission dominates the image. 
The surface brightness of the shells also varies substantially across the sample. 
While some novae exhibit bright, well-defined shells, others are considerably fainter, with low surface-brightness emission. 
In this sense, the signal-to-noise ratio of the observations of the low surface-brightness nova shells V1974\,Cyg and T\,Aur is insufficient to reliably investigate their spatially-resolved spectral properties, and even to extract one-dimensional spectra. 
These two sources are therefore excluded from the 3D analysis and spectral characterization, although they are retained in the sample for completeness.

\subsection{One-dimensional spectra}
\label{sec:Data_spec}

In order to assess the spectral properties of the nova shells in our sample, 1D spectra have been extracted from the data cubes for all but the faint objects V1974\,Cyg and T\,Aur. 

The 1D spectra with best signal-to-noise are presented in Fig.~\ref{Fig:spec_sample3} of Appendix~\ref{sec:ApendiceA}. Prior to extraction, the contribution from the central system  was removed whenever detected, so that the resulting spectra represent only the nebular emission.
In general the extracted spectra include only a limited number of strong nebular emission lines consistently detected across the sample. 
These are the H$\alpha$ and H$\beta$ emission lines, which are detected in most nova shells in the sample. 
The [O\,{\sc iii}]~$\lambda\lambda4959,5007$ and [N\,{\sc ii}]~$\lambda\lambda6548,6584$ nebular lines are also detected, but many nova shells lack emission in these lines.   

\begin{figure}
\centering
\includegraphics[width=\linewidth]{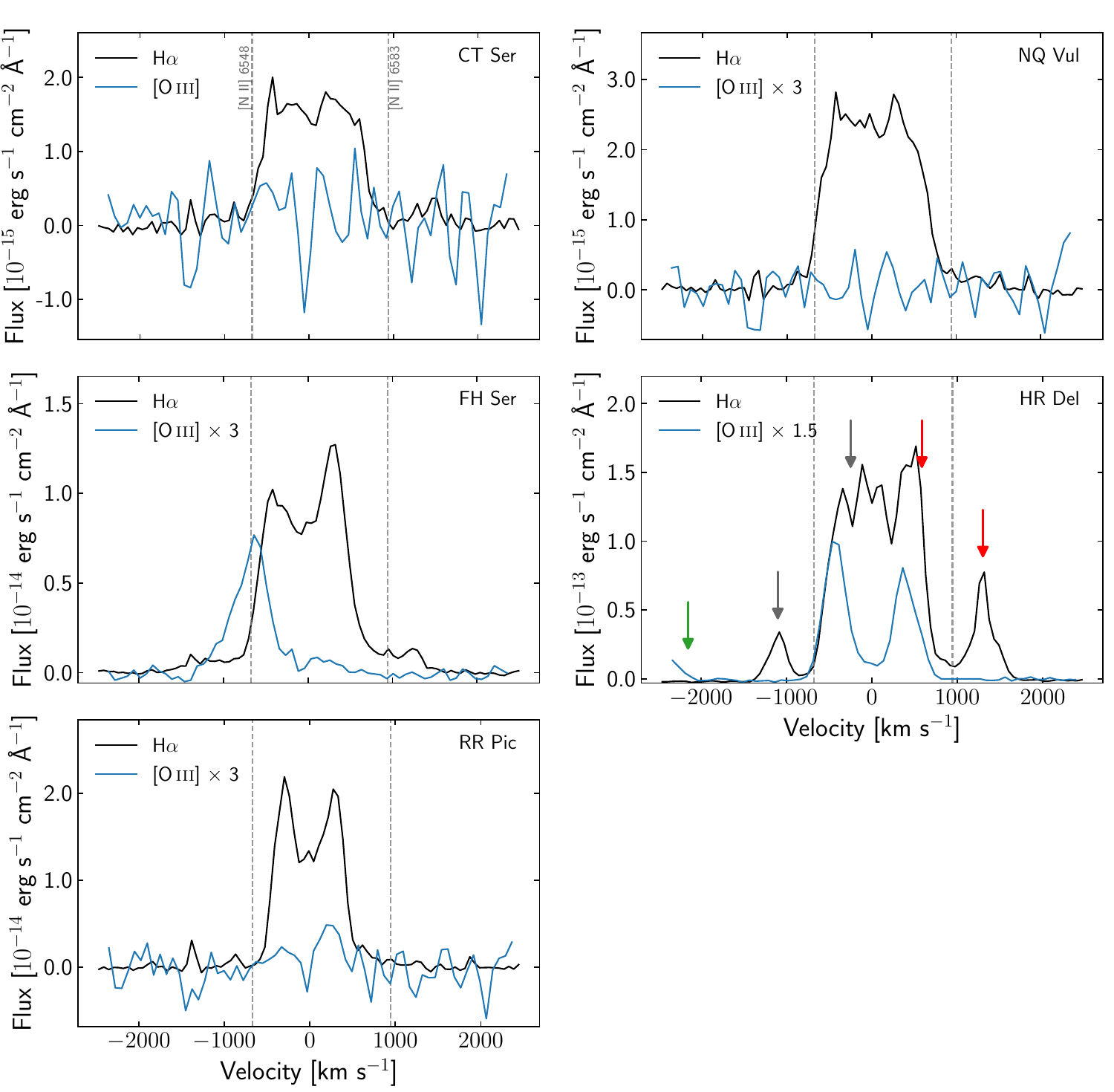}
\caption{
H$\alpha$ (black) and [O\,{\sc iii}]~$\lambda5007$ (blue) spectral profiles of a selected sample of nova remnants: CT\,Ser (top-left), NQ\,Vul (top-right), FH\,Ser (middle-left), HR\,Del (middle-right), and RR\,Pic (bottom-left).  
Gray dashed lines mark the expected velocities of the [N~{\sc ii}] $\lambda\lambda6548,6583$ lines relative to H$\alpha$. Gray and red arrows in the HR\,Del panel identify the high-velocity components of [N~{\sc ii}] $\lambda\lambda6548,6583$ respectively.
Only FH\,Ser and HR\,Del show noticeable emission in [O~{\sc iii}]. The rise at velocities bluewards in the [O III] spectral profile of HR Del (green arrow) corresponds to the red component of the [O~{\sc iii}]$\lambda4959$ emission line
}
\label{Fig:spec_sample}
\end{figure}

We extracted H$\alpha$ and [O~{\sc iii}]~$\lambda5007$ line profiles from the one-dimensional nebular spectra.  
Some examples are shown in Fig.~\ref{Fig:spec_sample}, while the rest of the sample are shown in Fig.~\ref{Fig:spec_sample2}. 
These profiles provide a direct view of the velocity structure of the emitting gas that will be later compared to the 3D visualizations provided by the IFS data. 
The profiles typically span a velocity range of $\pm800$ km~s$^{-1}$, reflecting the high expansion velocities of the ejecta. 
Note the contamination of the H$\alpha$ line by the [N\,{\sc ii}]~$\lambda\lambda6548,6584$ lines in some cases, as also noted in Tab.~\ref{tbl:nova.info}. 

The shape of the H$\alpha$ spectral profiles varies notably among the different sources.  

Their main shapes can be described as 
single-peaked (SP), double-peaked (DP), multi-peaked (MP), and flat-top (FT) spectral profiles. 
Additional secondary descriptors, such as asymmetric (a) or castellated (c) spectral profiles, the latter showing multiple small amplitude peaks, will also be considered.  
The classification of the H$\alpha$ spectral profiles is listed in the last column of Tab.~\ref{tbl:nova.info}.  

\subsection{3D visualization}
\label{sec:Data_3D}

The 2D images in Fig.~\ref{FigVibStab} have been used to describe the morphology of these nova shells, but they are actually projections onto the plane of the sky of emissivity-weighted 3D structures.  
This limitation is overcome by IFS observations, which provide simultaneous spatial and kinematic information to explore both the morphology and expansion of the ejecta of nova shells. 
Therefore, by combining the spatial distribution of the emission with the line-of-sight velocity information encoded in the emission-line profiles at each position, IFS data allow us to investigate the intrinsic 3D structure of nova shells beyond limited 2D projections.

Prior to constructing the 3D visualizations, the sky background and stellar continuum emissions were removed from each nova emission line. 
For this purpose, spectral channels adjacent to the nova emission line free of sky line emission were selected to construct a sky background and stellar continuum map. 
This was subsequently subtracted to every channel including the corresponding nova emission line.

To access easily to this information, position-position-velocity (PPV) visualizations of the nova shells were constructed using {\sc Mayavi} \citep{Mayavi2011} and 
tailored {\sc python} routines such as {\sc Astropy}\footnote{{\sc Astropy} is a Python package designed explicitly for Astronomy and maintained by the astronomical community \citep{astropy:2013, astropy:2018, astropy:2022}.} and the {\sc APLpy} tools, an open-source Python plotting package \citep{Aplpy2012}. 
The visualization pipeline follows the methodology presented by \citet{Santamaria2022b} and further developed by \citet{Guerrero+2025}. 

It allows the construction of PPV directly from the IFS data cubes. 

\begin{figure*}
   \centering
   \includegraphics[width=1.0\linewidth]{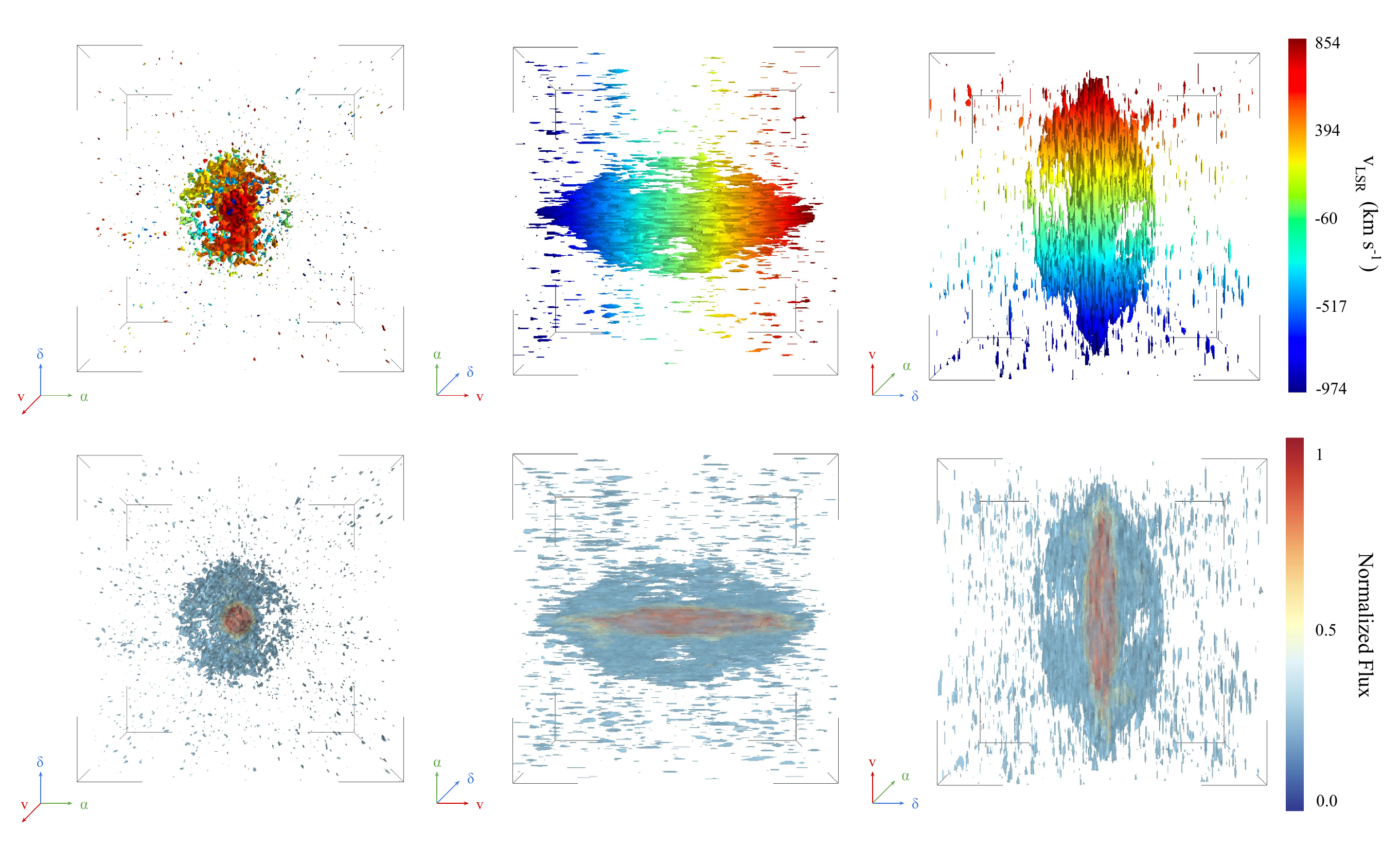}
      \caption{ H$\alpha$ PPV visualization of CT Ser. The upper panels show the radial velocity distribution, colour-coded according to the line-of-sight velocity, while the lower panels display the corresponding normalized flux distribution. From left to right, the panels present the projection along the observer’s line of sight (plane of the sky, $\alpha$--$\delta$), followed by the $\lambda$--$\alpha$ and $\delta$--$\lambda$ projections. 
              }
         \label{fig:CTSer}
\end{figure*}

\begin{figure*}
   \centering
   \includegraphics[width=1.0\linewidth]{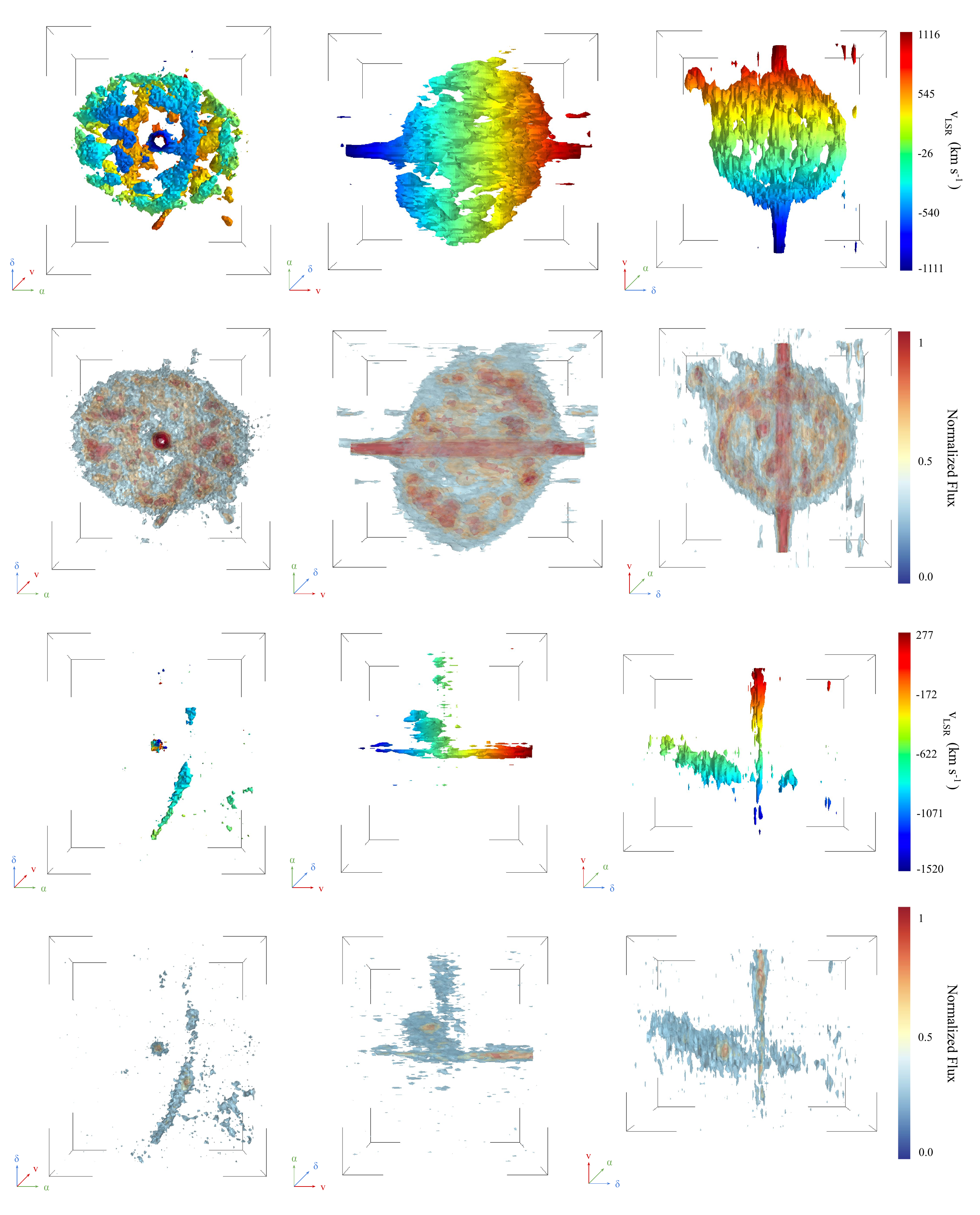}
      \caption{H$\alpha$ (top) and [O~{\sc iii}]$\lambda5007$ (bottom) PPV visualizations of FH Ser.}
      \label{FHSer}
\end{figure*}

\begin{figure*}
   \centering
   \includegraphics[width=1.0\linewidth]{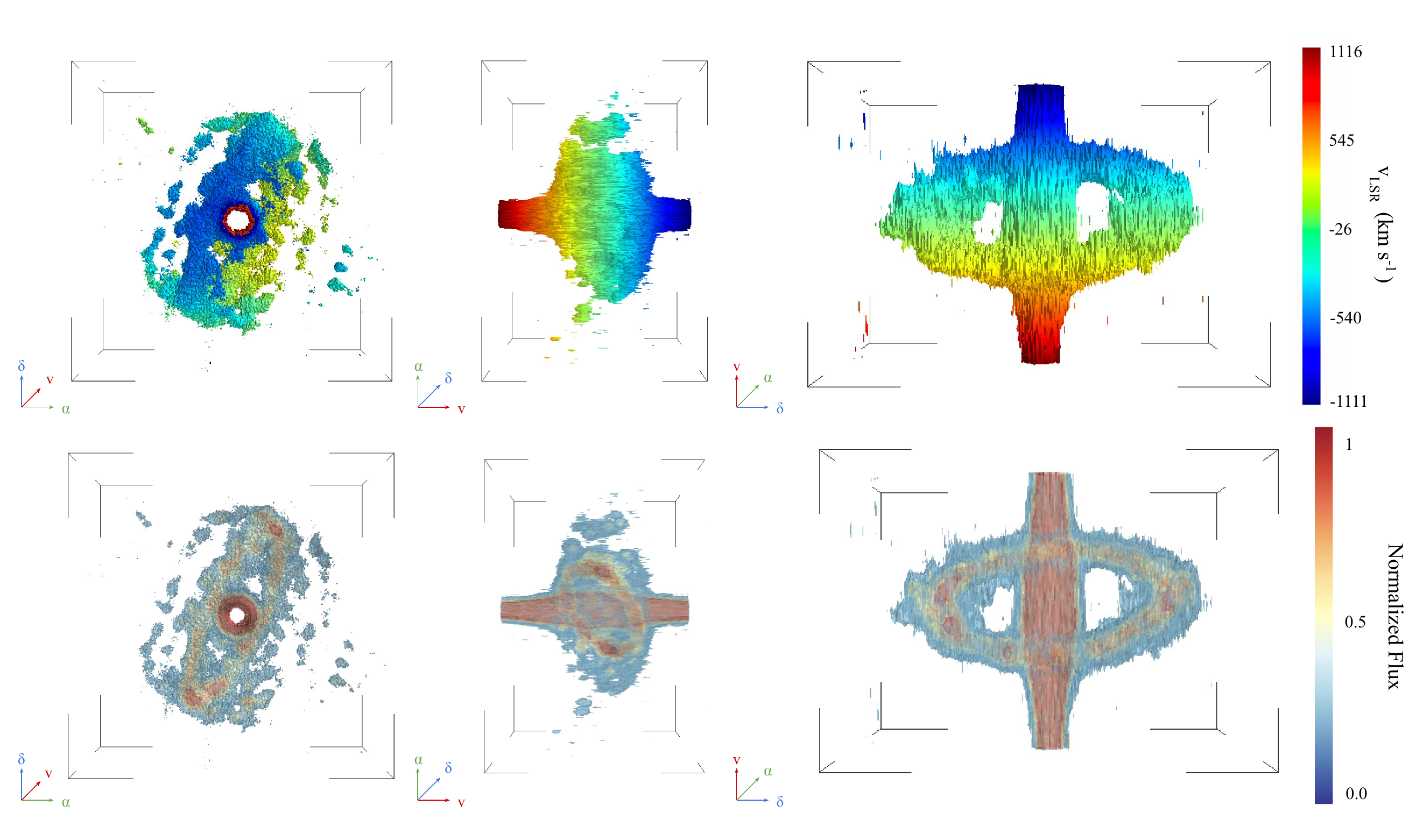}
      \caption{H$\alpha$ + [N~{\sc ii}] PPV visualization of RR Pic.
              }
         \label{RRpic}
\end{figure*}

\begin{figure*}
   \centering
   \includegraphics[width=1.0\linewidth]{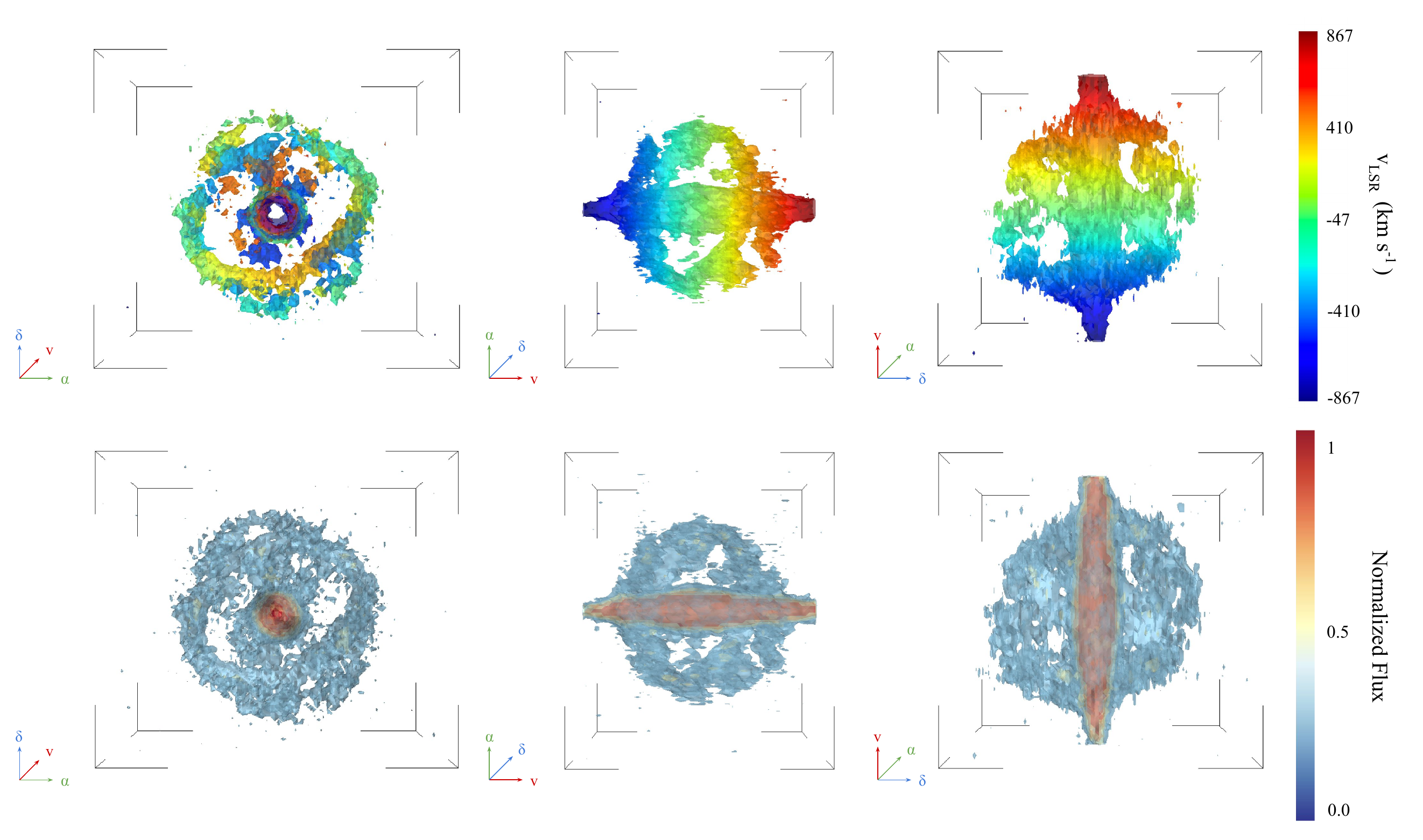}
      \caption{H$\alpha$ PPV visualization of NQ Vul.
              }
         \label{NQvul}
\end{figure*}

\begin{figure*}
    \centering
    \includegraphics[width=\textwidth]{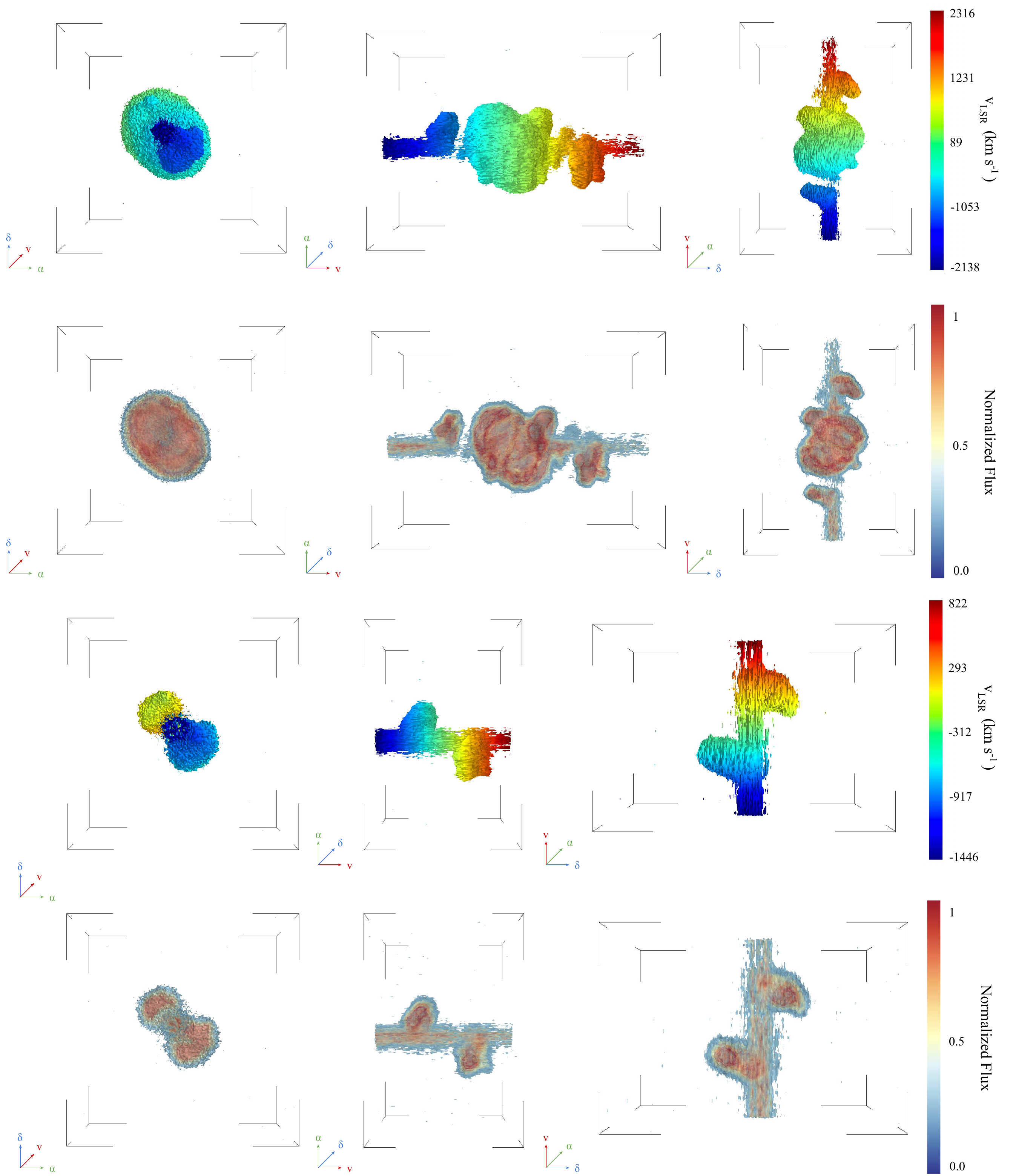}
      \caption{H$\alpha$ (top) and [O~{\sc iii}]$\lambda5007$ (bottom) PPV visualization of HR Del.}
    \label{HRDel}
\end{figure*}

Examples of representative nova shells are shown in Figs.~\ref{fig:CTSer} to \ref{HRDel}, with the complete set of PPV visualizations for the remaining objects in the sample being presented in Appendix~\ref{sec:ApendiceB}. 
In the following section, we present the results derived from these PPV visualizations. 

\section{Results}
\label{sec:results}

\subsection{Morphology versus Physical Structure}

The images of the nova shells shown in Fig.~\ref{FigVibStab} display a variety of morphological types, including approximately round or elliptical shells (e.g. CT\,Ser, BT\,Mon), ring-like structures (e.g. RR\,Pic, NQ\,Vul), bipolar (e.g. HR\,Del, V5668\,Sgr) and more irregular or asymmetric morphologies (e.g. Nova\,Cyg\,J210204). 
Their morphological classification, according to the ERBIAS (elliptical, round, bipolar, irregular, asymmetric, stellar) scheme developed for planetary nebulae \citep{Parker+2006} and adapted to nova shells by \citet{Santamaria2025a}, is also provided in column 8 of Tab.~\ref{tbl:nova.info}. 
As described there, secondary morphological features are assigned using the lower case ‘scxftme’ letters for (s)mooth, (c)lumpy, or mi(x)ed (smooth and clumpy) surface brightness, (f)ilaments, knots with (t)ails, (m)ultiple shells, and surface brightness (e)nhancement along the minor axis. 
As expected from the analysis of a larger sample of nova shells \citep{Santamaria2025a}, the sample is dominated by approximately round or mildly elliptical morphologies, which are 15 out of 19.

The morphology of a nova shell is the projection onto the plane of the sky of its emissivity-weighted 3D physical structure.  
As shown in IFS studies of nova shells \citep[e.g.,][]{Santamaria2022b,Guerrero+2025}, the examination of multiple projections of a nova shell and the comparison of the velocity and intensity distributions allow the characterization of their true 3D physical structure. 
Furthermore these PPV visualizations are particularly effective for identifying asymmetries, clumpiness, and departures from axisymmetry in nova shells that may not be evident in 2D projections alone, such as images or position-velocity diagrams extracted from long-slit spectroscopic observations.

The 3D physical structure of the nova remnants in the sample is compared to their morphology in the last column of Tab.~\ref{tbl:nova.info}.  
These have been determined from the PPV visualizations shown in Figs.~\ref{fig:CTSer} to \ref{HRDel} and Appendix~\ref{sec:ApendiceB}.  
When talking about 3D physical structures, R means a spherical (round) shell, E an ellipsoidal shell, and B an hour-glass structure.  
It is needed to add here an additional descriptor for (T)oroidal structures, as some nova shells actually consists mainly of emission along a wide belt. 
We note here that the equatorial emission enhancement denoted by the ``e'' secondary morphological attribute might hint at the presence of a tilted toroidal structure not fully unveiled in direct images.

\subsection{Individual Case Studies}

We present next a few examples of PPV visualizations of nova remnants.  
These are shown from less to higher degree of complexity and feature richness. 

One of the simplest sources is maybe CT\,Ser.  
The PPV visualizations of its H$\alpha$ line (Fig.~\ref{fig:CTSer}) reveal a mildly clumpy ellipsoidal shell.  
Its radial expansion velocity is much lower than the tangential expansion velocity, suggesting it is an oblate ellipsoid, although its distance and its uncertainty question this conclusion.  

Note that the morphology assigned by \citet{Santamaria2025a} was Rc (round with clumps). 
The H$\alpha$ spectral line profile of the nebular emission (shown in the top-left panel of Fig.~\ref{Fig:spec_sample}) can be described as top-flat, possibly with multiple narrow-components overimposed (FT-c), although at the level of the noise in the continuum emission. 
This is also consistent with the low level of clumpiness in the PPV visualizations. 

The PPV visualizations of the H$\alpha$ line of FH\,Ser (Fig.~\ref{FHSer}) are in this case indicative of an ellipsoidal shell notably tilted with the line of sight. 
Its radial expansion velocity is lower than the tangential expansion velocity, suggesting it is an oblate ellipsoid.  
However, the PPV visualizations of the [O~{\sc iii}] $\lambda5007$ emission line displays a filamentary structure that seems to be laying along the equator of the ellipsoidal shell seen in H$\alpha$.  
This same feature seems to be present in the [N~{\sc ii}] $\lambda6584$ emission line, but its relative weakness with respect to the H$\alpha$ emission hampers an unambiguous description. 
This equatorial component is described in the VLT VIMOS IFS investigation of this nova shell presented by \citet{Guerrero+2025}. 
The PPV visualizations thus confirm the morphological classification of FH\,Ser, elliptical with clumps, filaments and enhanced equatorial emission (Ecef).  
 
The H$\alpha$ spectral line profile (shown in the middle-left panel of Fig.~\ref{Fig:spec_sample}) shows an uneven double-peaked (DP) shape, with most emission confined within approximately $\pm800$ km s$^{-1}$ 
Meanwhile the [O~{\sc iii}]~$\lambda5007$ line is significantly weaker and confined to a narrower velocity range preferentially concentrated at the blue-shifted side, suggesting that the higher-ionization gas traces a particular kinematic component. 
This behavior is consistent with the PPV visualizations, where the [O~{\sc iii}] emission appears arranged in a linear or filamentary structure rather than forming a closed shell.

As noted in FH\,Ser, wherever the [O~{\sc iii}] emission line is detected, its spectral profile differs significantly from that of H$\alpha$, either in velocity extent, relative peak intensity or overall shape (see the middle-left panel of Fig.~\ref{Fig:spec_sample} and Fig.~\ref{Fig:spec_sample2} ).  
The [O~{\sc iii}] emission line profile of some novae appears confined to higher-velocity components, consistent with the structures revealed in their PPV visualizations.

A different case is that of RR~Pic. 
The PPV visualizations of the H$\alpha$ emission line in  Fig.~\ref{RRpic} shows the material distributed mainly on a well-defined toroidal structure, with no evidence for a shell-like structure. 
The orthogonal projections $\lambda-\alpha$ and $\lambda-\delta$ clearly show the velocity gradient within this toroidal structure, implying that it is tilted with respect to the line of sight. 
Meanwhile, the direct image in Fig.~\ref{FigVibStab} and left panels of Fig.~\ref{RRpic} reveal faint clumps aligned orthogonally to the toroidal structure.  
Thus, in this case, the 3D physical structure revises the morphological classification (elliptical with clumps with tails and enhancement along the minor axis) in the sense that the elliptical shape does not correspond to a tilted prolate ellipsoid, but to an equatorial toroidal structure. 
This interpretation is broadly consistent with the results presented by \citet{Celedon2024}, which also reported a strongly axisymmetric structure with equatorial enhancement in RR\,Pic, reinforcing the idea that its ejecta is dominated by a ring- or torus-like component.
Despite the overall different structure of RR\,Pic and FH\,Ser, their H$\alpha$ spectral profiles show both a clear double-peaked (DP) profile with small expansion velocity.

Somehow in between the shell-like structure of CT\,Ser or FH\,Ser and the toroidal RR\,Pic is the case of NQ~Vul.  
The PPV visualizations of its H$\alpha$ line (Fig.~\ref{NQvul}) reveal a largely symmetric shell that can be interpreted as a ring/torus with clumps along the symmetry axis, specially clear in the $\lambda-\alpha$ projection. 
The H$\alpha$ spectral profile is flat-top (top-right panel of Fig.~\ref{Fig:spec_sample}), with multiple faint peaks superimposed on the main emission component (FT-c), whereas the [O~{\sc iii}]~$\lambda5007$ line is not clearly detected.

Finally the PPV visualizations of HR\,Del (Fig.~\ref{HRDel}) provide one of the clearest examples of their great diagnostic power. 
The PPV visualizations of the H$\alpha$ and [N\,{\sc ii}] emission lines in the $\lambda-\alpha$ and $\lambda-\delta$ projections reveal an ellipsoidal shell in H$\alpha$, but an equatorial enhancement and two faster bipolar caps in [N~{\sc ii}].  
The physical structure of the material responsible for the [N~{\sc ii}] emission, which is contaminated by the H$\alpha$ line, is much clearly seen in the [O~{\sc iii}] $\lambda5007$ emission line. 
This complex morphology, previously modeled using high-resolution images and long-slit spectra \citep{HarmanOBrien2003,Santamaria2022a}, is unambiguously recovered in the 3D visualization, confirming the intrinsically aspherical nature of the ejecta, particularly in the nebular lines of [O~{\sc iii}] and [N~{\sc ii}]. 
The physical structure is much complex than the observed morphology, Ecft, which reflects the ellipsoidal shell in H$\alpha$ and the clumpy emission in [N~{\sc ii}]. 
Note that the discrepant physical structure displayed by the H$\alpha$ emission line and the nebular lines of [O~{\sc iii}] and [N~{\sc ii}] was also the case of FH\,Ser (Fig.~\ref{FHSer}) and would also be the case whenever these different lines are detected (Fig.~\ref{Fig:spec_sample3}).  
As expected the H$\alpha$+[N~{\sc ii}] and [O~{\sc iii}] spectral profiles of HR\,Del in the middle-right panel of Fig.~\ref{Fig:spec_sample} are notably different. 
The former can be described as FT-c, resulting from the blended emissions of the H$\alpha$ ellipsoidal shell and the [N~{\sc ii}] bipolar lobes, while the latter displays two coherent velocity components arising from the bipolar lobes.

\section{Discussion}
\label{sec:Discussion}

The discussion will first focus on the PPV visualizations of the objects in the H$\alpha$ emission line, which provides the most homogeneous tracer across the sample. 
V1974\,Cyg and T\,Aur, which are too faint, are excluded, whereas J210204\,Cyg will be investigated through its [N~{\sc ii}] emission line.

A number of 3D physical structures are identified among the 17 resolved nova remnants in Tab.~\ref{tbl:nova.info} with definite 3D structure in H$\alpha$, as well as [N~{\sc ii}] in the case of J210204\,Cyg, including ellipsoidal/spherical shells, ring-like structures, and bipolar features. 
Approximately 60$\%$ of the novae exhibit predominant ellipsoidal 3D structures, with only one (RR\,Cha) with radial and tangential velocities consistent with a spherical shell, that would be a $\simeq$6\% of the sample. 
Another 30$\%$ show toroidal/ring-like or bipolar structures or both, and 18$\%$ display irregular structures. 
Some present mixed or ambiguous 3D structures.  

In contrast to the statistical results on nova remnants morphology reported by \citet{Santamaria2025a}, where elliptical morphologies completely dominate the sample and bipolar shapes were infrequent, the 3D PPV diagrams reveals a comparatively smaller fraction of ellipsoidal shells and a larger proportion of toroidal or bipolar structures. 
Indeed objects previously classified as elliptical in 2D imaging reveal a different 3D structure in PPV diagrams. 
Such are the cases of V5668~Sgr, which shows a clear bipolar structure in the PPV reconstruction, and V1229\,Aql, which exhibits a more irregular morphology than previously inferred.

[O~{\sc iii}] emission is detected in only five of the 17 nova remnants with available (MUSE or WEAVE) sensitive observations covering this spectral range. 
Among these, two object exhibit a clear bipolar morphology in [O~{\sc iii}], while the remaining three show irregular or strongly asymmetric structures. Although limited by small-number statistics, this result suggests that [O~{\sc iii}] preferentially traces dynamically or physically distinct components of the ejecta than H$\alpha$, reinforcing the need for multi-line, 3D studies to fully characterize nova shells.

The sky and radial expansion velocities listed in Tab.~\ref{tbl:nova.info} are compared in Fig.~\ref{fig:vsky_vrad}. 

The comparison of these velocity components needs to consider their different nature and methodology used to determine them. 
The tangential expansion velocity is derived from the angular radius of the shell, the time since eruption, and the adopted distance, resulting in an average expansion velocity. 
On the other hand, the radial expansion velocity is obtained directly from spectroscopic measurements, thus corresponding to an instantaneous velocity. 
Since nova shells remain in free expansion until the time they disperse into the ISM \citep{Santamaria2020}, this behavior makes the sky expansion velocity, which is a value averaged in time, to coincide with the present day sky expansion velocity. 
This allows its comparison with the radial expansion velocity, which is an instantaneous value.

Most sources in Fig.~\ref{fig:vsky_vrad} are within the gray-shaded area, which represents values in the range $1.18 > V_{\mathrm{sky}}/V_{\mathrm{rad}} > 1/1.18 = 0.85$. 
This gray-shaded area has been selected because it corresponds to the typical aspect ratio of ellipsoidal nova shells, which differs by 0.18 times the radius of a sphere according to the statistical study of the images of 45 resolved nova shells presented by \citet{Santamaria2025a}. 
Therefore the novae within the gray-shaded area of Fig.~\ref{fig:vsky_vrad} have velocity ratios consistent with the aspect ratios derived from the statistical study of resolved nova remnants. 
There are, however, a few sources that have significantly large values of $V_{\mathrm{sky}}/V_{\mathrm{rad}}$. 
This is reasonable for BT\,Mon, RR\,Pic, and FH\,Ser, which have toroidal physical structures or equatorial enhancements that for small inclination angles will result in values of $V_{\mathrm{sky}}$ larger than $V_{\mathrm{rad}}$. 
For the other cases, we need to examine the systematics that affect the determination of $V_{\rm rad}$ and $V_{\rm sky}$.

The determination of $V_{\rm rad}$ and its comparison with $V_{\rm sky}$ can be affected by different systematics, including the use of different emission lines, contamination by other emission lines, and dynamical effects.  
The first two have been avoided by deriving $V_{\rm rad}$ from the same emission line used to measure the angular radius of the nova, and by a reliable identification in 2D velocity maps of the approaching and receding velocity components. 
Dynamical effects, on the other hand, may result from the deceleration of the expansion after sweeping ISM material up to a mass comparable to that of the nova ejecta.  
This will increase the $V_\mathrm{sky}$ to $V_\mathrm{rad}$ ratio, as the first is an instantaneous value and the second an average on time.  
Indeed the most discrepant sources in Fig.~\ref{fig:vsky_vrad} (V351\,Pup, CT\,Ser and XX\,Tau) have the largest physical sizes, with radii in the range 0.10 to 0.15 pc (Tab.~\ref{tbl:nova.info}). 
We note, however, that at least CT\,Ser and XX\,Tau, would be located far away from the Galactic Plane, at 3.0 and 1.5 kpc, respectively, where the ISM density is expected to be very low and the swept up mass would be negligible.  

Meanwhile systematics in the determination of the distance and angular size uncertainty directly propagate into $V_{\rm sky}$. 
\citet{Santamaria2025a} noted that the distances to nova remnants derived from Gaia data seem to be well determined up to 3 kpc, but they are systematically larger than those derived by \citet{Schaefer2022} beyond this distance. 
Indeed the discrepant sources in Fig.~\ref{fig:vsky_vrad}, namely V1425\,Aql, V351\,Pup, CT\,Ser, and XX\,Tau, have Gaia distances above 3 kpc (5.7, 12.0, 4.0, and 7.5 kpc, respectively). 
Their Gaia distances are thus prone to have been overestimated, which is considered the most likely reason of their anomalously high $V_\mathrm{sky}$ to $V_\mathrm{rad}$ ratio. 

If instead they were assumed to have to lay on the grey-shaded area of Fig.~\ref{fig:vsky_vrad}, i.e.\ $0.85 \leq V_\mathrm{sky}/V_\mathrm{rad} \leq 1.18$, their distances need to be corrected to $4.8\pm0.8$ kpc for V351\,Pup, $2.3\pm0.4$ kpc for CT\,Ser, and $3.3\pm0.8$ kpc for XX\,Tau. 
Otherwise the small angular size of V1425\,Aql, which is barely resolved, might have additionally contributed to overestimate its angular size and thus its sky velocity.

The 3D view offered by the PPV diagrams provides the means to interpret the shape of the spectral profiles of nova remnants (last column of Tab.~\ref{tbl:nova.info}), which is the only available information in young, spatially unresolved sources.  
Double-peaked (DP) spectral profiles of H$\alpha$ are indicative of expanding shells (XX\,Tau), but also toroidal structures (RR\,Pic). 
The most common spectral profiles are those characterized by multiple-peaks (MP) of large amplitude (V1425\,Aql) or by a top-flat shape, typically overimposed with uneven multiple small amplitude peaks (V351\,Pup), the so-called ``castellated'' profiles (FT-c).  
These correspond to sources with multiple structural components, including the equatorial waist/ring and lobes of bipolar sources (V5668\,Sgr), or expanding shells filled with emission (CT\,Ser) or with multiple clumps located at different radial velocities with respect to the observer (CP\,Pup). 
Therefore the comparison between the 3D structure and spectral profile classifications in Tab.~\ref{tbl:nova.info} suggests that there is not an univocal correspondence between them.

 \section{Summary}
\label{sec:summary}

We have presented a 3D morphological analysis of a sample of 19 nova shells based on IFS observations obtained from archival data and new observations. By combining spatially resolved spectroscopy with kinematic information, we constructed PPV visualizations that allow the intrinsic structure of the nova ejecta to be explored beyond the limitations imposed by two-dimensional projections such as images or long-slit spectra.

The PPV visualizations provide a coherent view of the intrinsic structure of the ejecta. While the sample is dominated by apparently round or mildly elliptical morphologies in projection, the 3D analysis reveals a wider diversity of physical structures, including not only ellipsoidal shells, but also toroidal/ring-like and bipolar components. 
In several cases, the 3D structure significantly refines or revises the morphological classification inferred from imaging alone.  
On the other hand, the comparison between tangential and radial velocities confirms that nova shells are intrinsically only mildly aspherical.

The comparison between H$\alpha$ and [O\,{\sc iii}] emission highlights that different lines trace distinct regions of the ejecta. 
The [O\,{\sc iii}] emission line is often associated with more localized structures.

Individual objects illustrate this diversity. CT~Ser represents a relatively simple, mildly clumpy ellipsoidal shell, whereas FH~Ser shows a more complex structure with equatorial enhancements and filamentary components traced by higher-ionization lines. RR~Pic is dominated by a well-defined toroidal structure, while NQ~Vul appears as an intermediate case between shell-like and toroidal configurations. HR~Del exhibits one of the most complex structures, with different emission lines tracing distinct physical components, including an ellipsoidal shell, bipolar lobes and equatorial enhancements.

Overall, this work demonstrates that PPV visualizations based on IFS provide a powerful and physically motivated framework for characterizing nova shells. 
It confirms the concern that direct images might be affected by geometry and projection effects.
The catalog presented here establishes a reference for future studies of nova ejecta and offers new insights into the role of geometry, kinematics, and ionization in shaping the remnants of nova eruptions.

This study represents the first step in a broader investigation of nova shells based on integral-field spectroscopy. Future papers in this series will focus on detailed analyses of individual objects, including Nova Cyg and V5668 Sgr, combining their spatially resolved morphology, kinematics, and three-dimensional structure to further constrain the physical processes governing the shaping and evolution of nova ejecta.

\section{Data availability}

Most of MUSE observations used in this study are publicly available through the ESO Science Archive. The proprietary observations obtained by the authors are available from the corresponding author upon reasonable request.

\begin{acknowledgements}
JBRG and MAG acknowledge financial support from the Severo Ochoa grant CEX2021-001131-S funded by MCIN/AEI/10.13039/501100011033 and the MCIU grant PID2024-162229NB-I00 cofunded with FEDER funds.  
JBRG and JAT thanks support from the UNAM DGAPA PAPIIT project IN~102324. 
This article is based on observations obtained at GTC and WHT, and archival data from VLT MUSE. Based on observations made with the William Herschel Telescope operated on the island of La Palma by the Isaac Newton Group of Telescopes in the Spanish Observatorio del Roque de los Muchachos of the Instituto de Astrofísica de Canarias (WEAVE proposal WS2024A2-012). Based on observations collected at the European Organisation for Astronomical Research in the Southern Hemisphere under ESO. Based on observations made with the Gran Telescopio Canarias (GTC), installed at the Spanish Observatorio del Roque de los Muchachos of the Instituto de Astrofísica de Canarias, on the island of La Palma. This article is based on observations made in the Observatorios de Canarias del IAC with the WHT operated on the island of La Palma by the 
Instituto de Astrofísica de Canarias in the Observatorio del Roque de los Muchachos. 
This work has made extensive use of NASA's Astrophysics Data System.

\end{acknowledgements}

\bibliographystyle{aa}
\bibliography{reference}

\begin{appendix}
   \section{One-dimensional spectra}
   \label{sec:ApendiceA}

\begin{figure*}
   \centering
   \includegraphics[width=\linewidth]{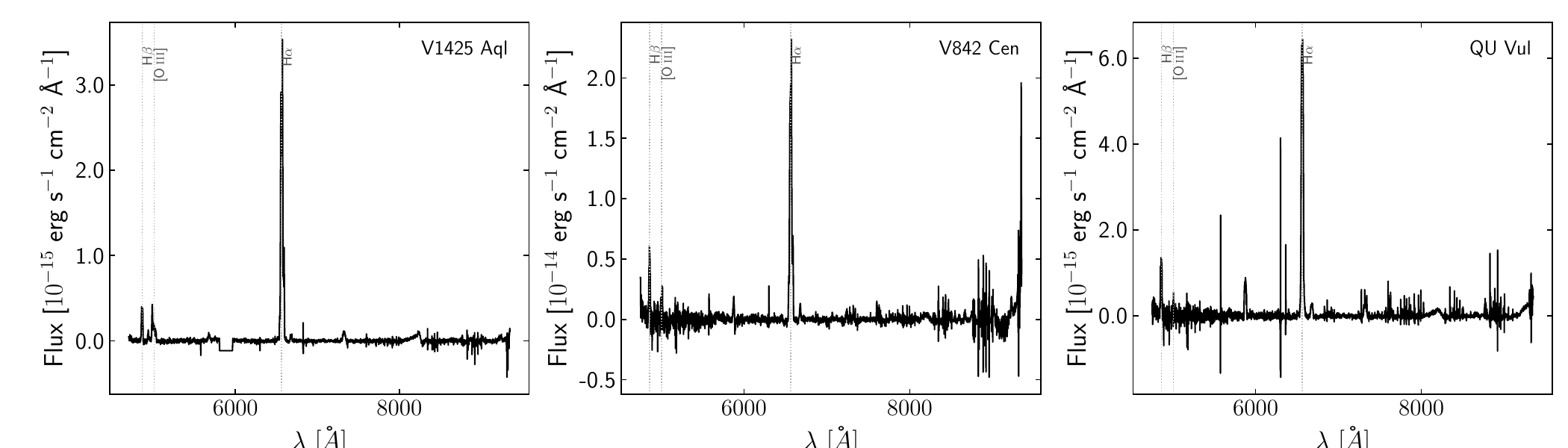}
   \caption{One-dimensional spectra of V1425~Aql (left), V842~Cen (middle) and QU~Vul (right).}
   \label{Fig:spec_sample3}
\end{figure*}

Fig.~\ref{Fig:spec_sample3} shows the full one-dimensional nebular spectra (after subtraction of the stellar contribution) of V1425~Aql, V842~Cen, and QU~Vul, which are the nova remnants with the best signal-to-noise ratio spectra. 
In all three objects, the strongest detected emission feature is H$\alpha$, while H$\beta$ and [O~{\sc iii}]~$\lambda5007$ are also detected with varying relative strengths. 

\begin{figure*}
\centering
\includegraphics[width=\linewidth]{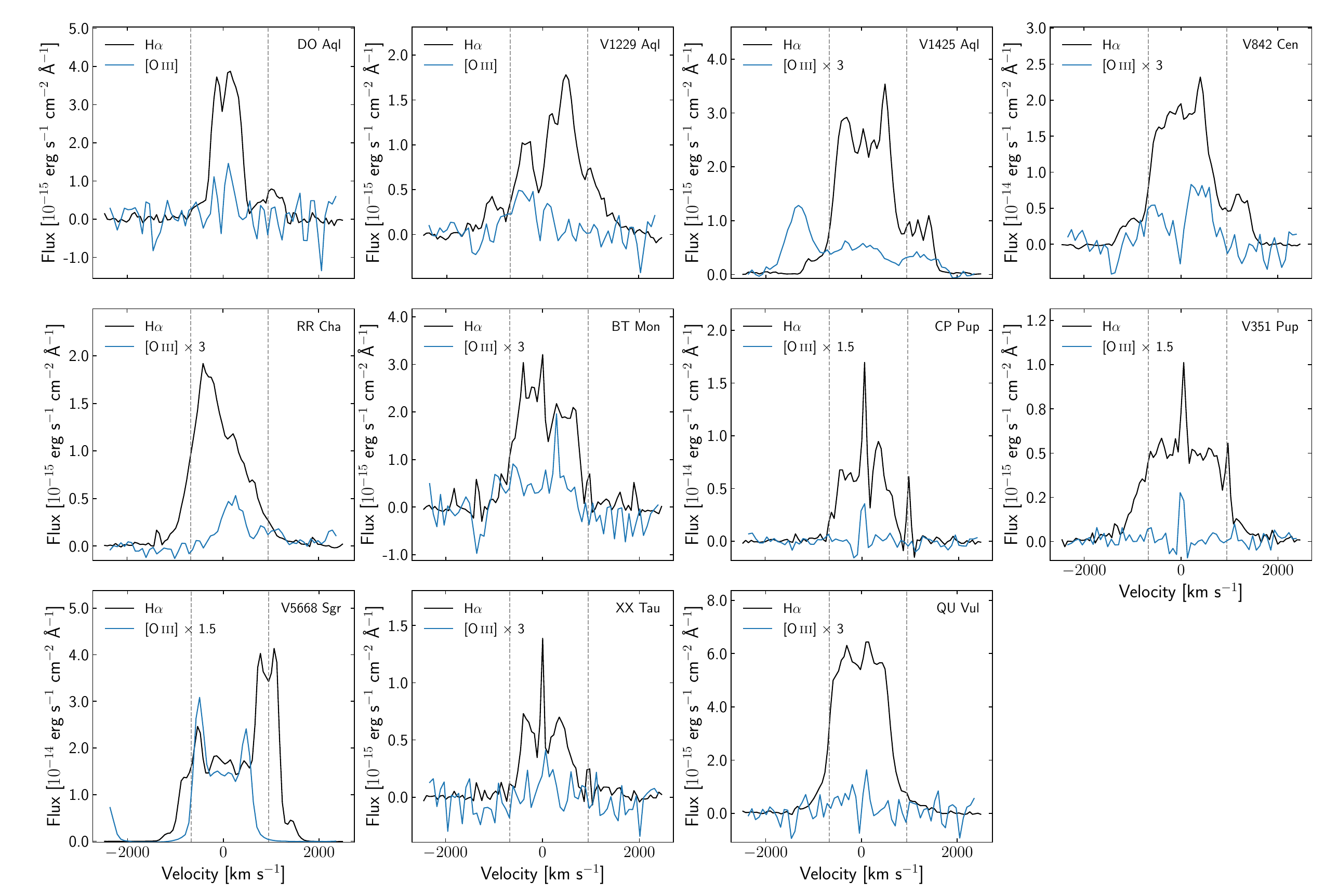}
\caption{
Spectral profiles of the H$\alpha$ (black line) and [O\,{\sc iii}]~$\lambda5007$ (blue line) emission lines of 
DO\,Aql, V1229\,Aql, V1425\,Aql, V842\,Cen (top), 
RR\,Cha, BT\,Mon,CP\,Pup, V351\,Pup (middle), V5688\,Sgr, XX\,Tau and QU\,Vul (bottom). Gray dashed lines mark the expected velocities of the [N~{\sc ii}] $\lambda\lambda6548,6583$ lines relative to H$\alpha$}
   \label{Fig:spec_sample2}
\end{figure*}

Fig.~\ref{Fig:spec_sample2} presents the extracted velocity profiles of H$\alpha$ and [O~{\sc iii}]~$\lambda5007$ for the remaining novae in the sample. The profiles display a wide variety of shapes, including flat-top, double-peaked, multi-component, and asymmetric structures, listed in Table~\ref{tbl:nova.info}. 

\section{PPV visualizations}
\label{sec:ApendiceB}
\begin{figure*}
   \centering
   \includegraphics[width=1.0\linewidth]{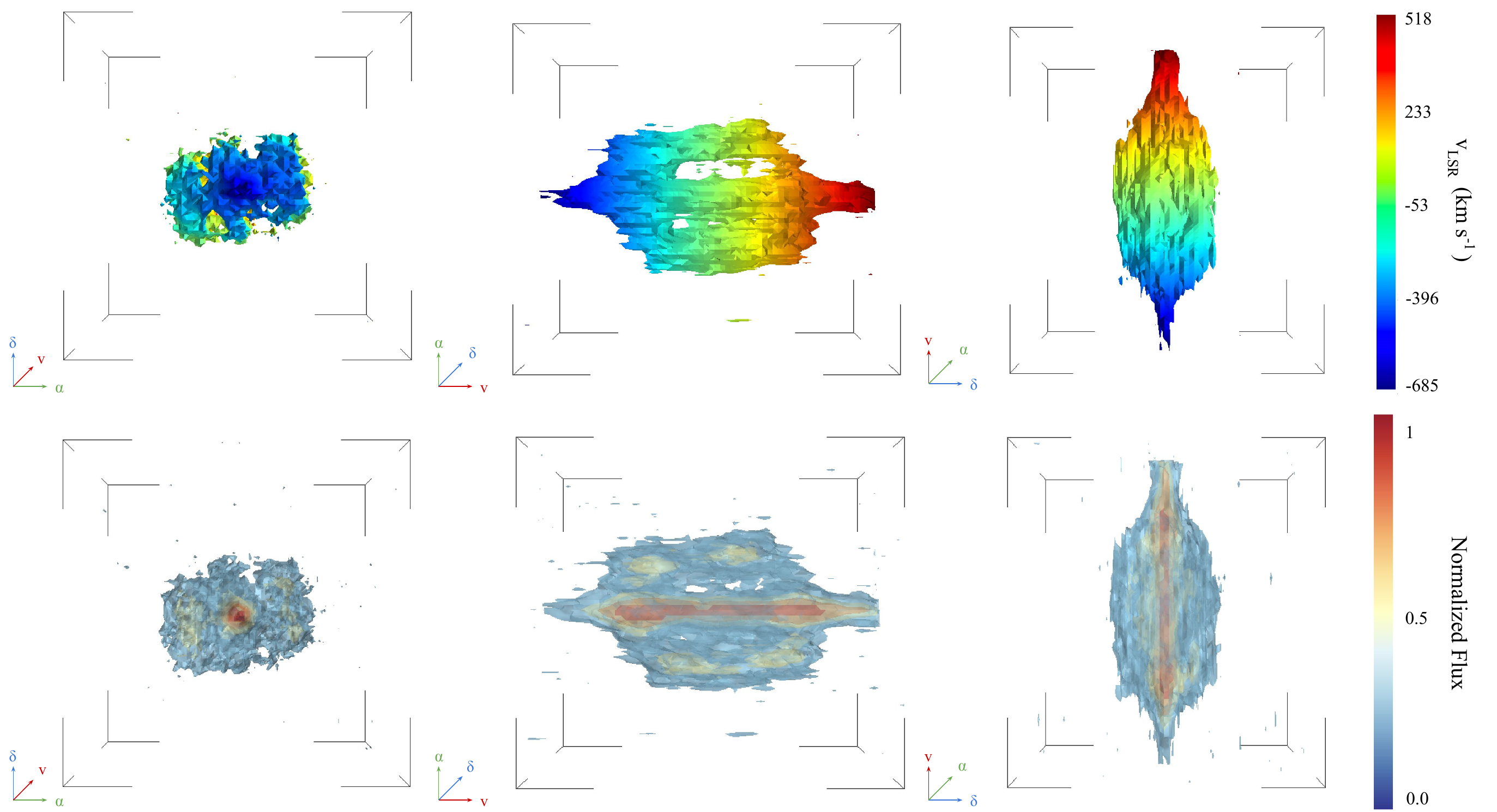}
      \caption{H$\alpha$ PPV visualization of DO Aql.
              }
         \label{DoAql}
\end{figure*}

\begin{figure*}
   \centering
   \includegraphics[width=1.0\linewidth]{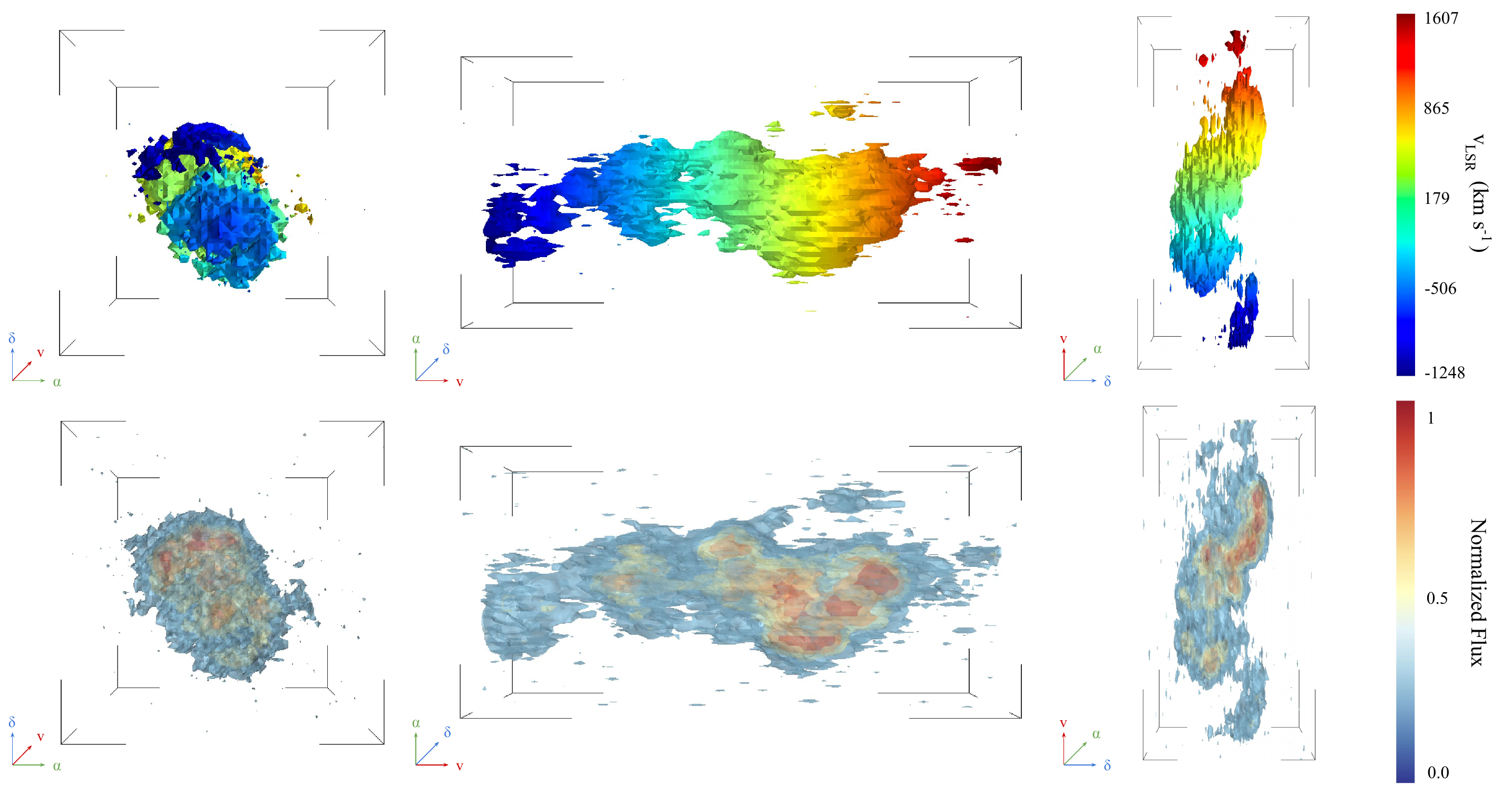}
      \caption{H$\alpha$ PPV visualization of V1229 Aql.
              }
         \label{V1229Aql}
\end{figure*}

\begin{figure*}[p] 
  \centering
  \begin{subfigure}{\textwidth}
    \centering
    \includegraphics[width=\textwidth]{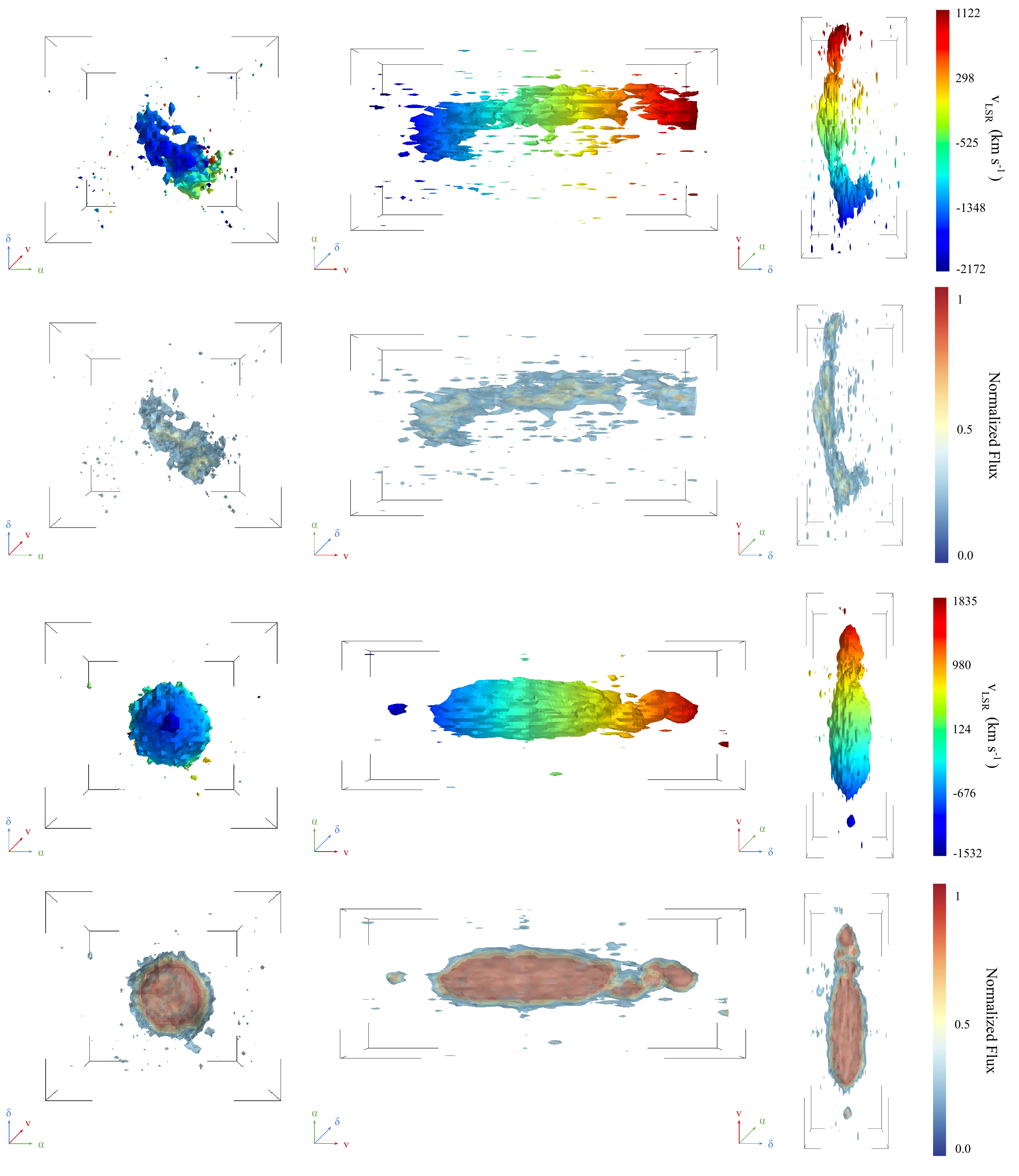}
    \label{fig:V1425Aql_a}
  \end{subfigure}
  \caption{[O~{\sc iii}]$\lambda5007$ (top) and H$\alpha$ (bottom) PPV visualizations of V1425 Aql.}
  \label{fig:V1425Aql}
\end{figure*}

\begin{figure*}
   \centering
   \includegraphics[width=1.0\linewidth]{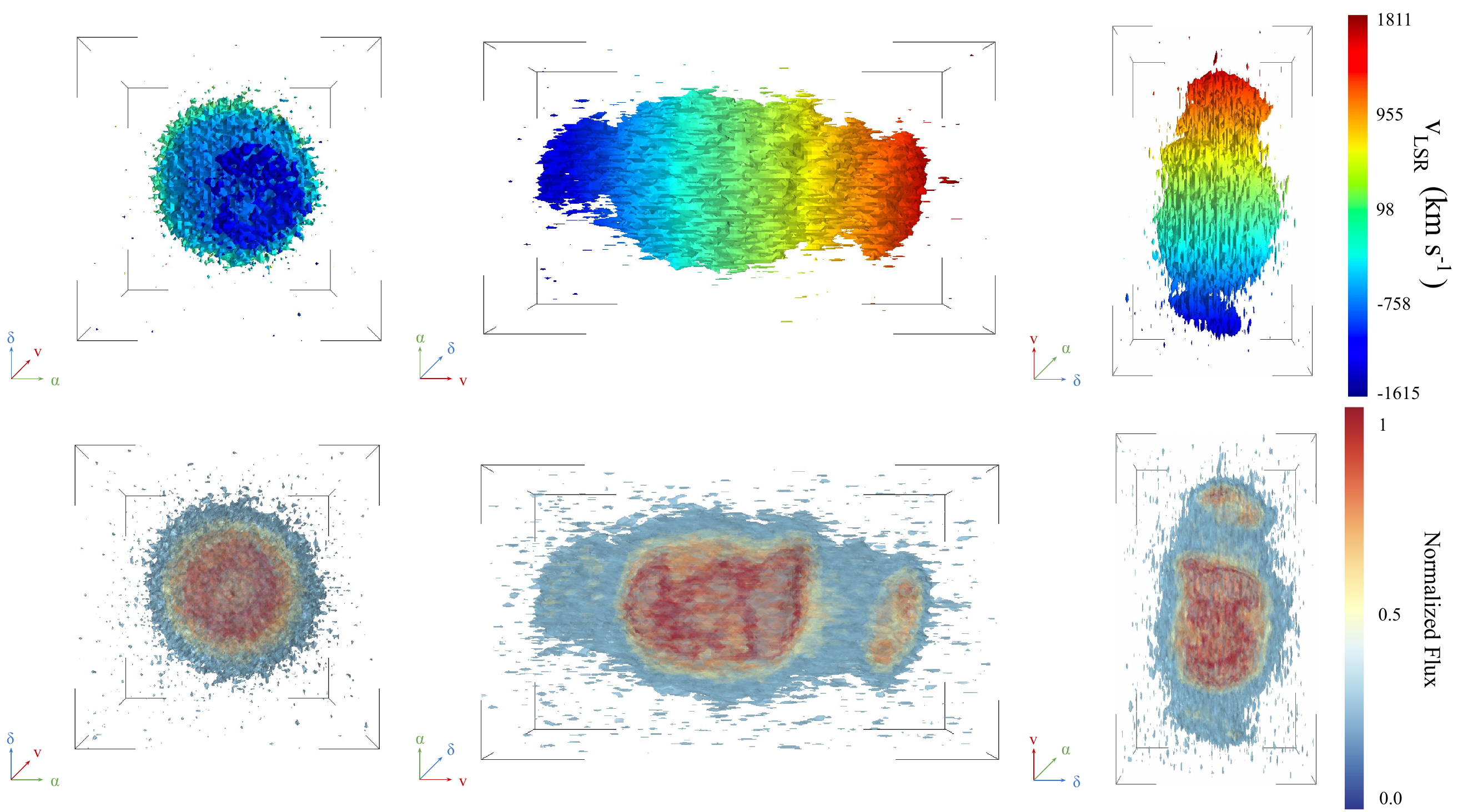}
      \caption{H$\alpha$ PPV visualization of V842 Cen.
              }
         \label{V842Cen}
\end{figure*}

\begin{figure*}
   \centering
   \includegraphics[width=1.0\linewidth]{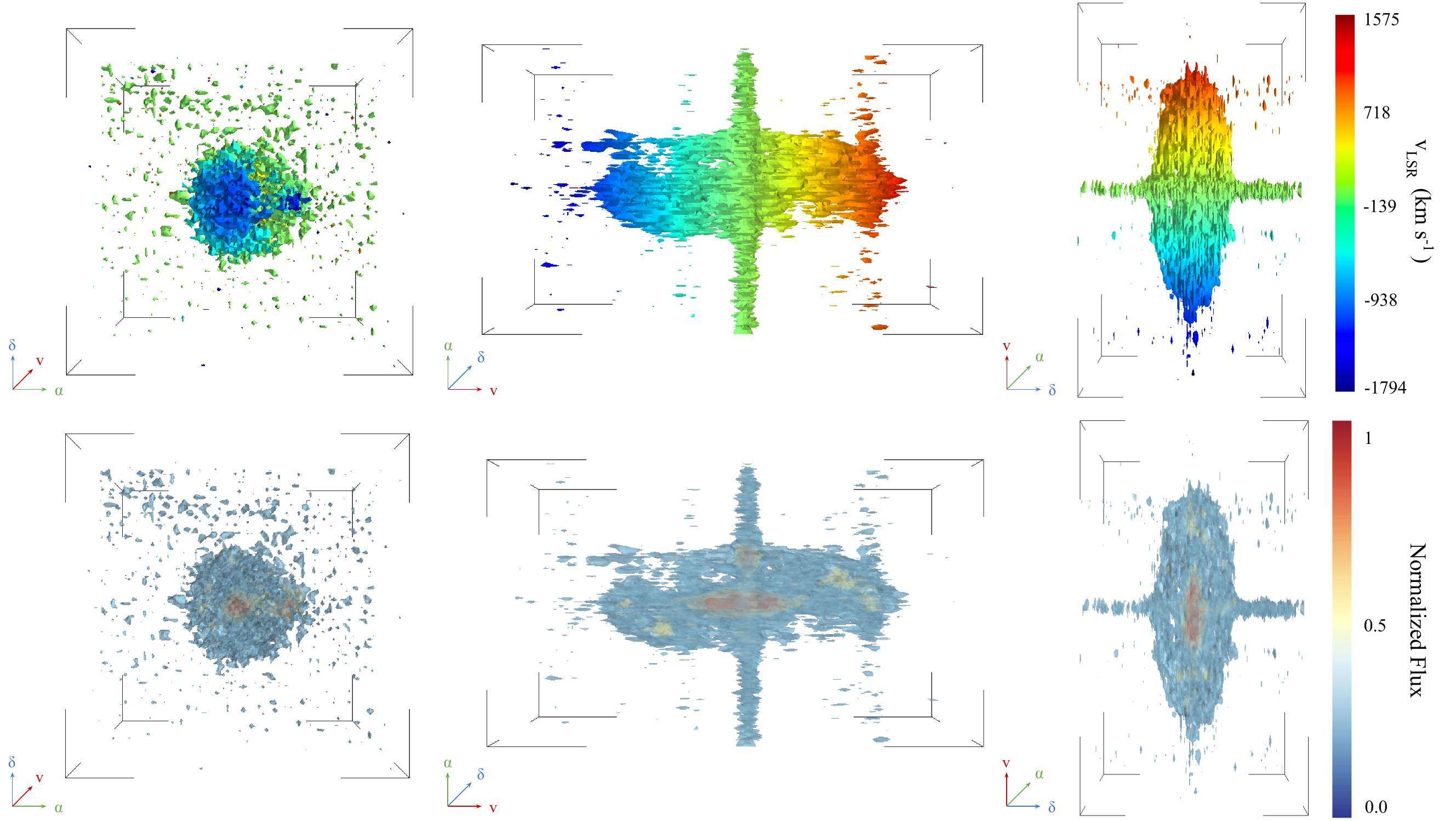}
      \caption{H$\alpha$ PPV visualization of V351 Pup.
              }
         \label{V351Pup}
\end{figure*}

\begin{figure*}
   \centering
   \includegraphics[width=1.0\linewidth]{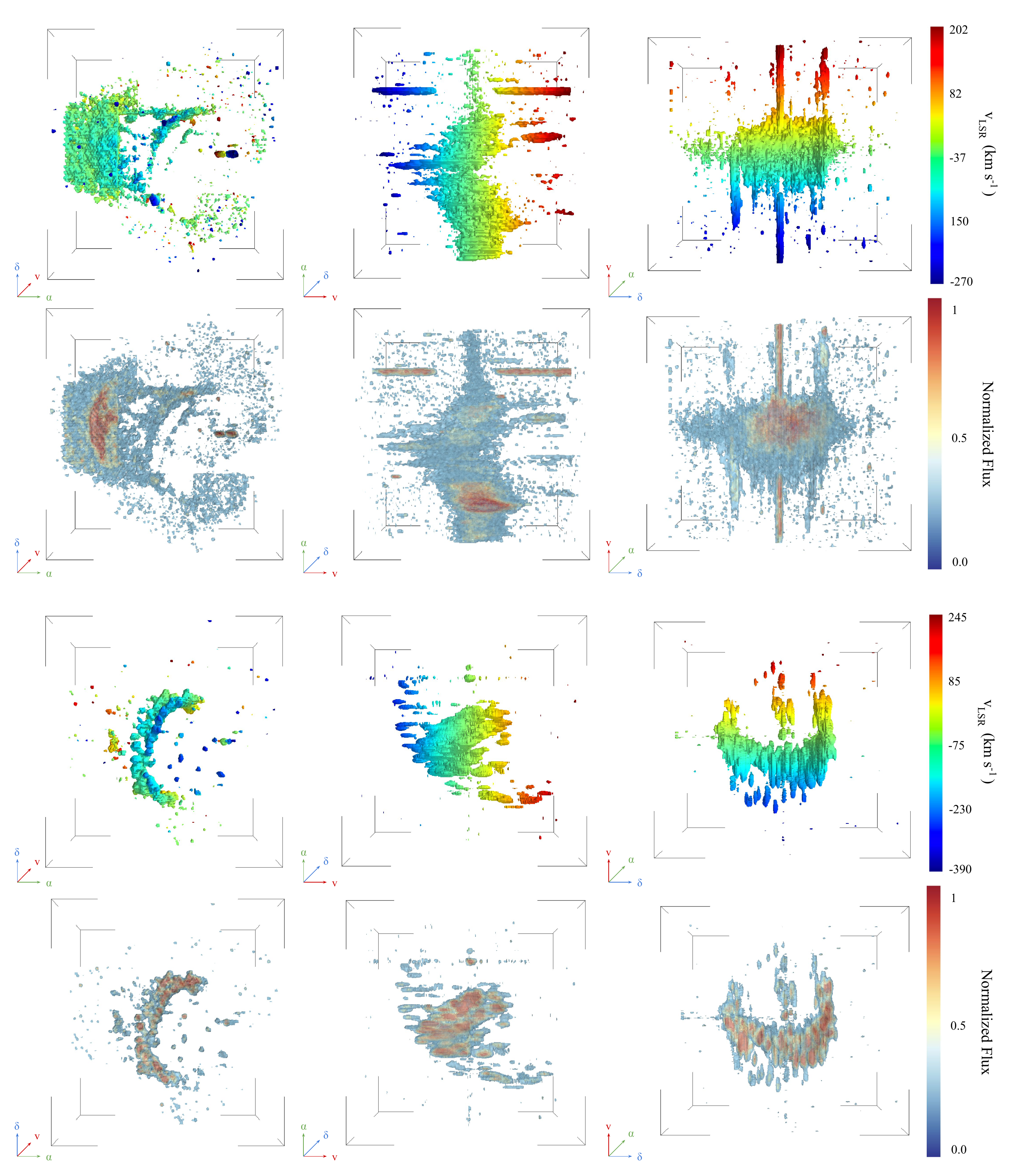}
      \caption{[O~{\sc iii}]$\lambda5007$ (top) and [N~{\sc ii}]$\lambda6584$ PPV visualization of J210204 Cyg.
              }
         \label{J210204}
\end{figure*}

\begin{figure*}
   \centering
   \includegraphics[width=1.0\linewidth]{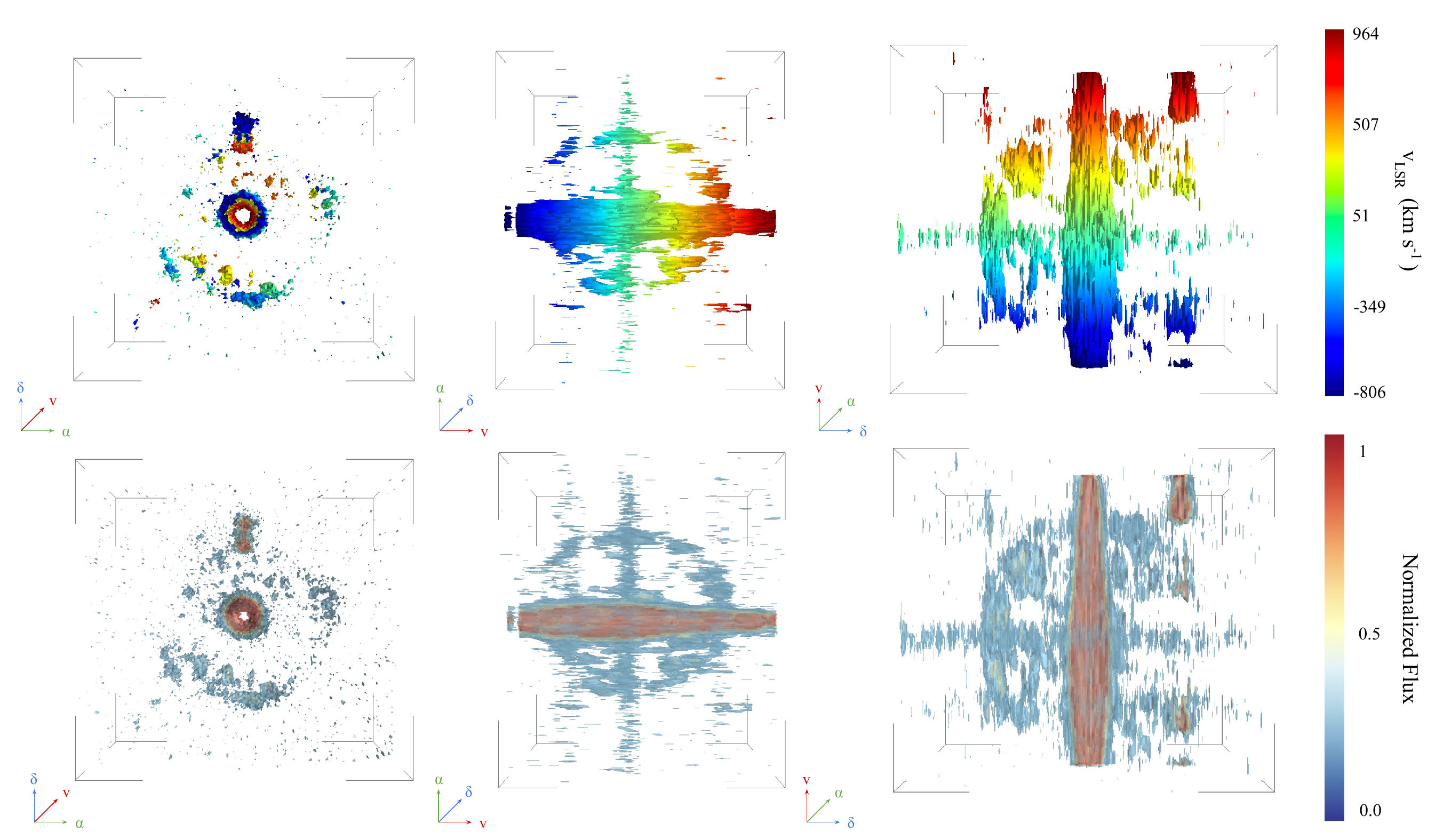}
      \caption{H$\alpha$ PPV visualization of BT Mon.
              }
         \label{BTMon}
\end{figure*}

\begin{figure*}
   \centering
   \includegraphics[width=1.0\linewidth]{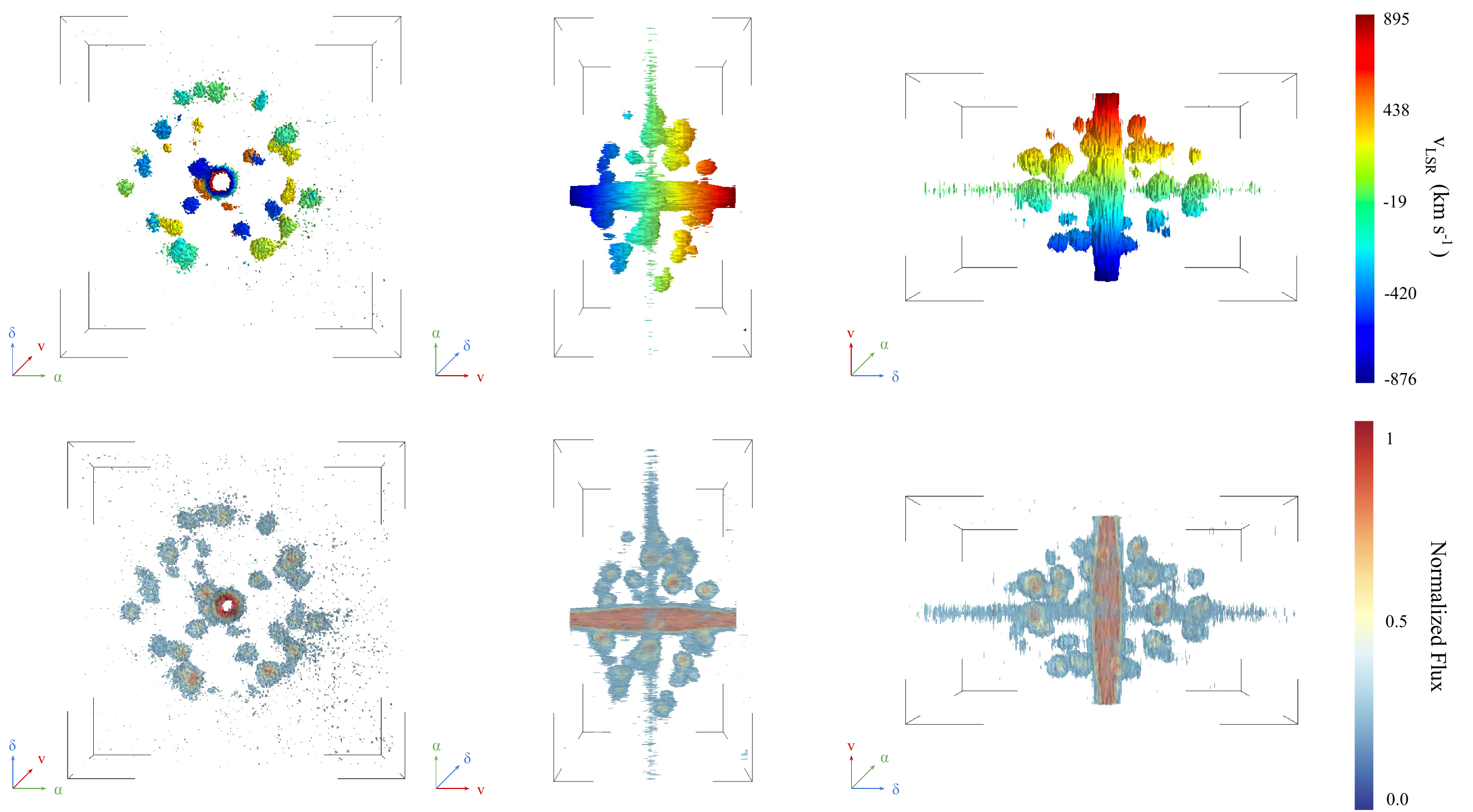}
      \caption{H$\alpha$ PPV visualization of CP Pup.
              }
         \label{CPpup}
\end{figure*}

\begin{figure*}
   \centering
   \includegraphics[width=1.0\linewidth]{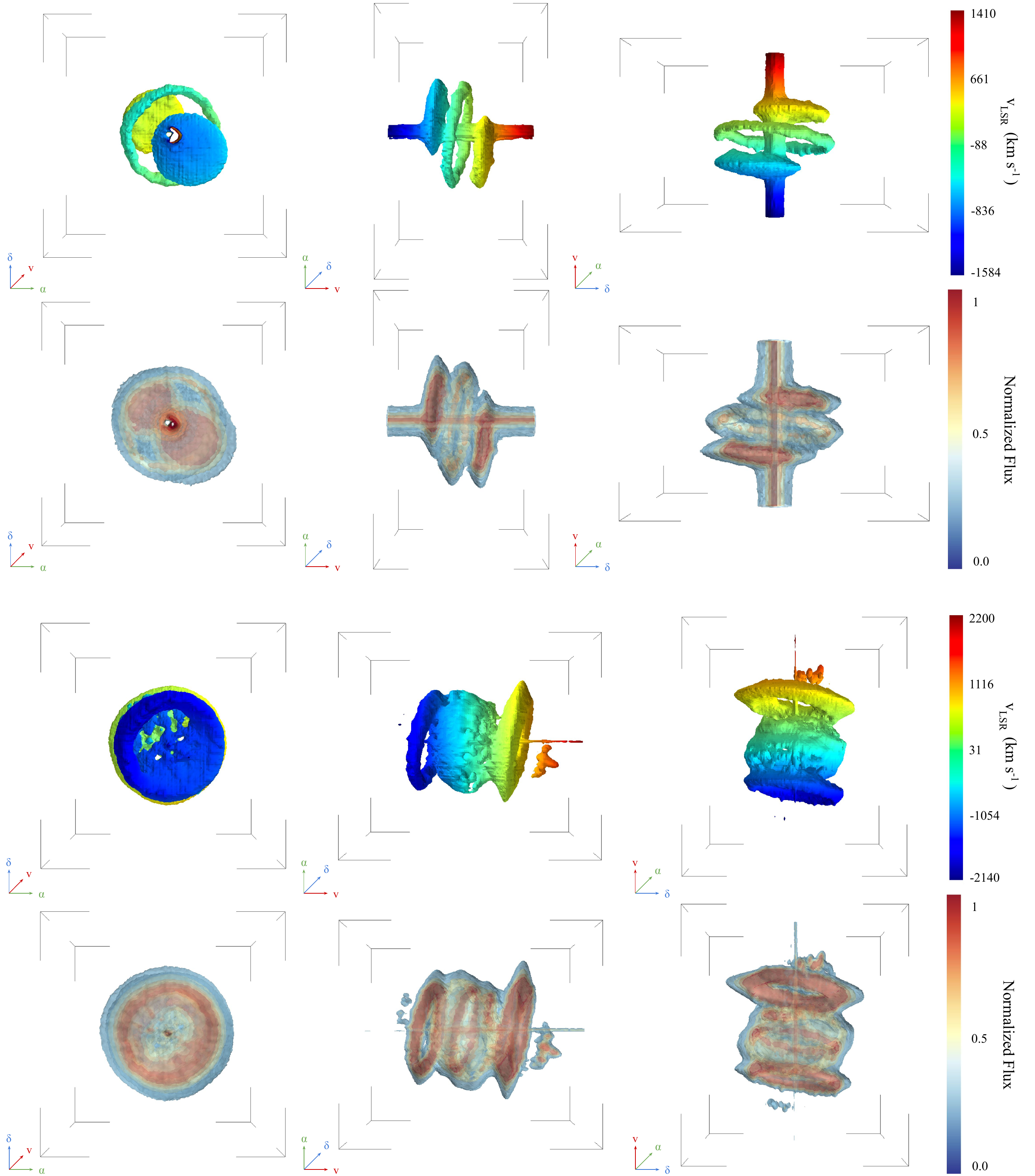}
      \caption{[O~{\sc iii}]$\lambda5007$ (top) and H$\alpha$+[N~{\sc ii}] (bottom) PPV visualizations of V5668 Sgr.
              }
         \label{V5668Sgr}
\end{figure*}

\begin{figure*}
   \centering
   \includegraphics[width=1.0\linewidth]{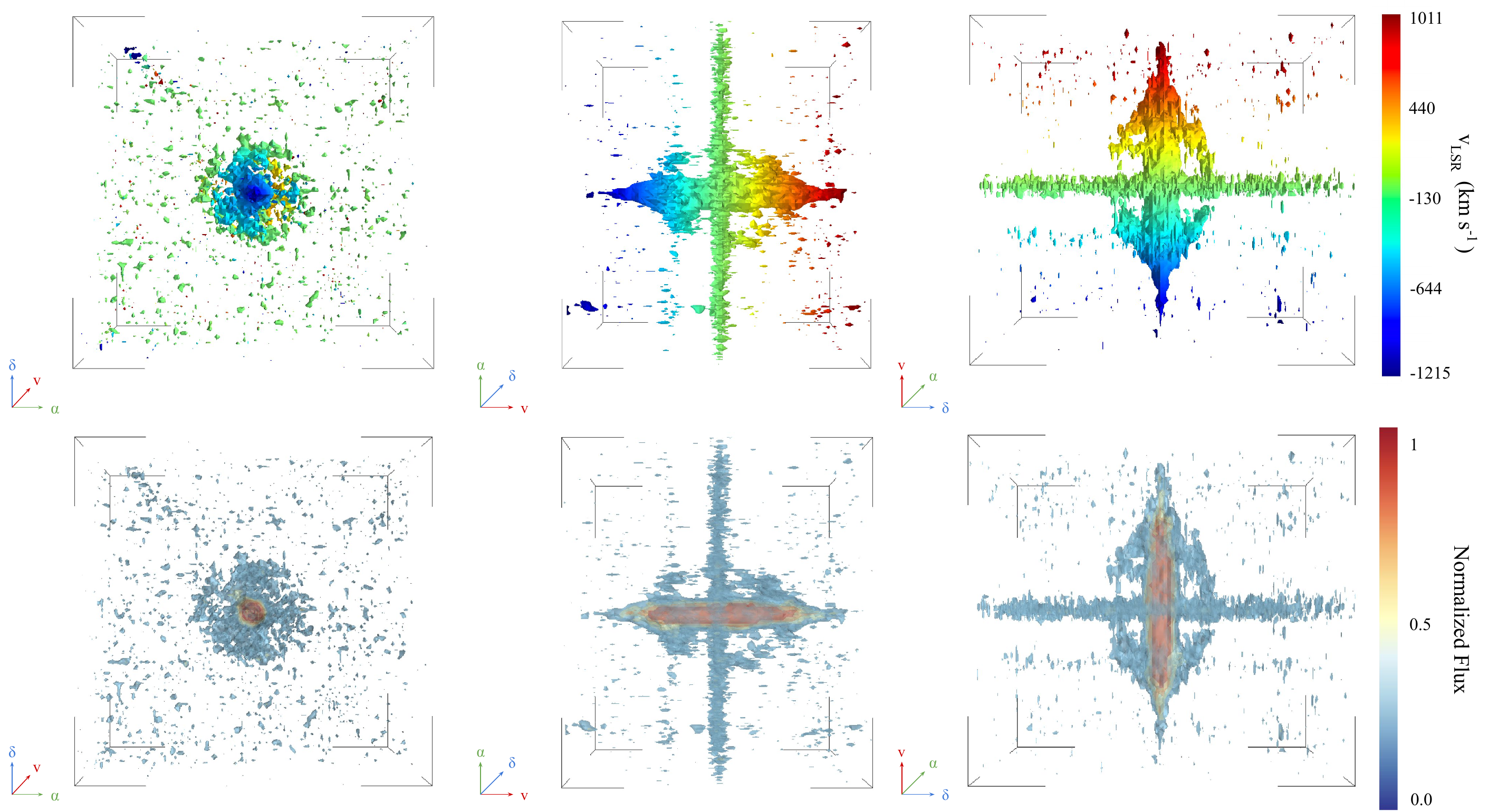}
      \caption{H$\alpha$ PPV visualization of XX Tau.               }
         \label{XXTau}
\end{figure*}

\begin{figure*}
   \centering
   \includegraphics[width=1.0\linewidth]{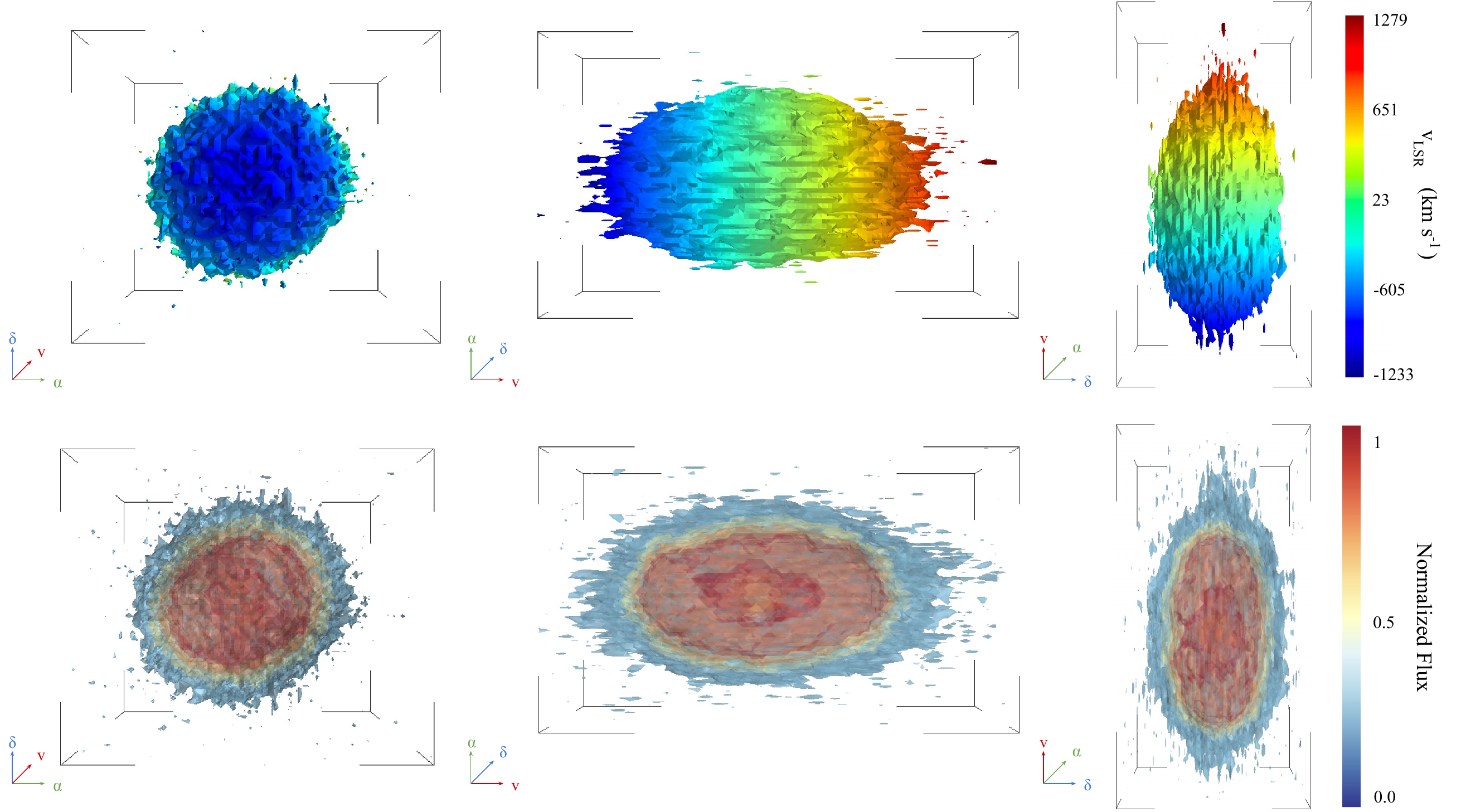}
      \caption{H$\alpha$ PPV visualization of QU\,Vul.
              }
         \label{Quvul}
\end{figure*}

\begin{figure*}[p] 
  \centering
  \begin{subfigure}{\textwidth}
    \centering
    \includegraphics[width=\textwidth]{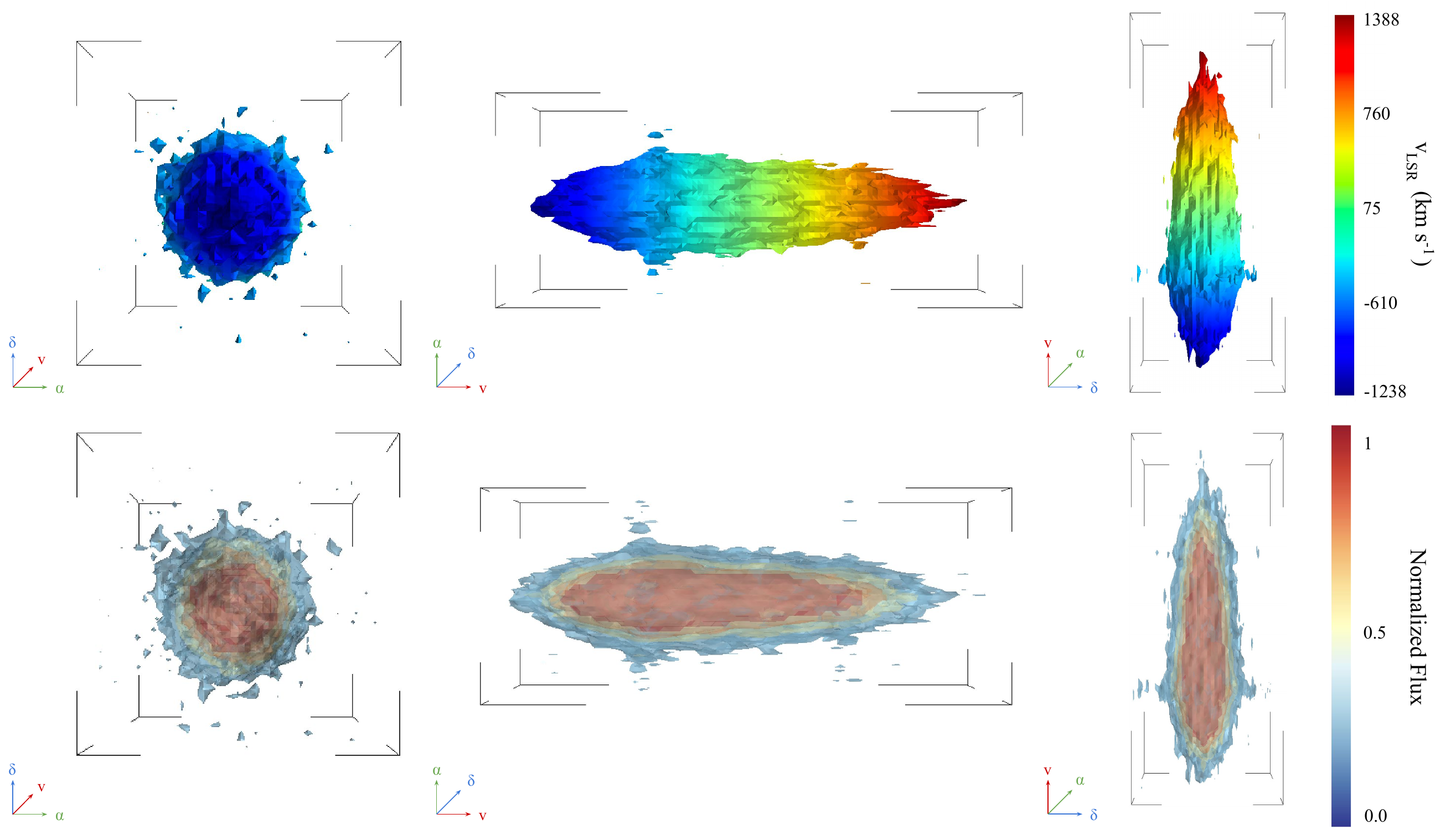}
    \label{fig:RRcha_a}
  \end{subfigure}
  \caption{H$\alpha$ PPV visualization of RR Cha}
  \label{fig:RRCha}
\end{figure*}

Figs.~\ref{DoAql} to \ref{fig:RRCha} present the PPV visualizations of the nova shells included in the catalog but not shown in the main body of the paper. 
These figures complement the representative examples discussed in Sect.~\ref{sec:Discussion} and provide a homogeneous view of the full sample.
While they are not discussed individually in the main text, they form an integral part of the catalog and provide the basis for future comparative studies of nova ejecta.

For each nova, the PPV visualization is constructed following the same methodology described in Sect.~\ref{sec:Data_3D}. The upper panels display the radial velocity distribution, colour-coded according to the line-of-sight velocity, while the lower panels show the corresponding intensity (flux) distribution. From left to right, the panels present the projection along the observer’s line of sight (i.e. the plane of the sky), followed by orthogonal projections in which the velocity axis replaces one of the spatial coordinates.


\end{appendix}

\end{document}